\documentclass[aps,pre,reprint,groupedaddress]{revtex4-2}
\usepackage{graphicx}
\usepackage{amsfonts}
\usepackage{amsmath}
\usepackage{bm}
\usepackage{mathtools}
\usepackage{newtxmath}
\usepackage{fix-cm}
\usepackage[export]{adjustbox}
\usepackage{color}

\newcommand{\Rey}{Re}
\newcommand{\JP}[1]{{\color{black}  #1}}
\newcommand{\AC}[1]{{\color{black}  #1}}

\def\lline{\vrule width14pt height2.5pt depth -2pt}
\definecolor{rea}{rgb}{0.12156862745098039, 0.4666666666666667, 0.7058823529411765}
\definecolor{reb}{rgb}{1.0, 0.4980392156862745, 0.05490196078431372}
\definecolor{rec}{rgb}{0.17254901960784313, 0.6274509803921569, 0.17254901960784313}
\definecolor{red}{rgb}{0.8392156862745098, 0.15294117647058825, 0.156862745098039}
\definecolor{ree}{rgb}{0.5803921568627451, 0.403921568627451, 0.7411764705882353}
\definecolor{ref}{rgb}{0.8901960784313725, 0.4666666666666667, 0.7607843137254902}
\definecolor{reg}{rgb}{0.5490196078431373, 0.33725490196078434, 0.29411764705882354}
\definecolor{reh}{rgb}{0.4980392156862745, 0.4980392156862745, 0.4980392156862745}
\definecolor{rei}{rgb}{0.7372549019607844, 0.7411764705882353, 0.13333333333333333}

\begin{document}

%\preprint{}

% \title{Estimating the dimension of the chaotic attractor in} two-dimensional Kolmogorov flow}
\title{Characterizing the Reynolds number dependence of the chaotic attractor in two-dimensional turbulence with dimension-minimizing autoencoders}


\author{Andrew Cleary}
\email[]{andrew.cleary@ed.ac.uk \\ Present address: Department of Mechanical Engineering, Johns Hopkins University, Baltimore, MD 21218, USA}
%\thanks{}
\affiliation{School of Mathematics \& Maxwell Institute for Mathematical Sciences, University of Edinburgh, Edinburgh, EH9 3FD, UK}

\author{Jacob Page}
\email[]{jacob.page@ed.ac.uk}
%\thanks{}
\affiliation{School of Mathematics \& Maxwell Institute for Mathematical Sciences, University of Edinburgh, Edinburgh, EH9 3FD, UK}


\date{\today}

\begin{abstract}

Deep autoencoder neural networks can generate highly accurate, low-order representations of turbulence. 
We design a new family of autoencoders which are a combination of a `dense-block' encoder-decoder structure (Page et al, \emph{J. Fluid Mech.} \textbf{991}, 2024), an `implicit rank minimization' series of linear layers acting on the embeddings (Zeng et al, \emph{Mach. Learn. Sci. Tech.} \textbf{5}, 2024) and a full discrete+continuous symmetry reduction. 
These models are applied to two-dimensional turbulence in Kolmogorov flow for a range of Reynolds numbers $25 \leq Re \leq 400$, and used to estimate the dimension of the chaotic attractor, $d_{\mathcal A}(Re)$.
We find that the dimension scales like $\sim Re^{1/3}$ — much weaker than known bounds on the \emph{global} attractor which grow like $Re^{4/3}$.
In addition, two-dimensional maps of the latent space in our models reveal a rich structure not seen in previous studies, including multiple classes of high-dissipation events at lower $Re$ which guide bursting trajectories. 
We visualize the embeddings of large numbers of ``turbulent’’ unstable periodic orbits, which the model indicates are distinct (in terms of features) from any flow snapshot in a large turbulent dataset, suggesting their dynamical irrelevance. This is in sharp contrast to their appearance in more traditional low-dimensional projections, in which they appear to lie within the turbulent attractor. 
% [[Suggested re-framing to come: Given large dimensionality of global attractor, question as to how many less DOFs are actually required to represent the physical/strange attractor actually sampled by the turbulence.
% We are seeking a scaling of \emph{attractor} dimension, which we must take care to distinguish from the \emph{global attractor} dimension. All the hallmarks of a strange attractor so maybe we just risk this terminology. 
% ]]}
\end{abstract}

%\keywords{}

\maketitle

\section{Introduction}
\label{sec:intro}

% intro outline

% 1) intro to dissipative pde and collapse onto inertial manifold, brief motivation for estimating dimensionality, statement of accomplishment of paper

Nonlinear partial differential equations (PDEs), such as the Navier–Stokes equations that govern turbulent fluid motion, are fundamental in modeling complex dynamics and often exhibit spatiotemporal chaos in parameter regimes of practical interest. 
Although the state spaces of PDEs are formally infinite-dimensional, it is argued that solutions of the Navier-Stokes equations collapse onto a finite-dimensional inertial manifold, denoted here by $\mathcal{M}$, due to the dissipative action of viscosity \cite{Hopf1948, temam1997infinite, Foias1988, TITI1990}.
% dissipative PDEs collapse onto an invariant finite-dimensional manifold, denoted by $\mathcal{M}$ \cite{Hopf1948, temam1997infinite}.
% This is formalized by the notion of an inertial manifold, a smooth finite-dimensional manifold containing the global attractor of the governing equation, which attracts all orbits at an exponential rate \cite{Foias1988, TITI1990}.
\AC{The inertial manifold is a smooth, finite-dimensional manifold, within which the \emph{global attractor} (which may be a complicated fractal set) is embedded \cite{Foias1988, TITI1990}.}
\AC{The global attractor $\mathcal A_G$ is defined as the set of points in phase space that can be reached in arbitrarily long time from all initial conditions \citep{Doering_Gibbon_1995}.}
\AC{Two familiar properties of $\mathcal A_G$ are (i) it attracts the time forward map of all initial conditions $\bm u_0$, such that $\text{dist}(\bm f^t(\bm u_0), \mathcal{A}_G) \to 0$ as $t \to \infty$, and (ii) \JP{it is an invariant set.}} %all solutions within $\mathcal A_G$ remain in $\mathcal A_G$, such that $\bm f^t(\bm u_0) \in \mathcal A_G$ for $\bm u_0 \in \mathcal A_G$.}
Although the existence of inertial manifolds for the Navier-Stokes equations has not been proven \cite{TEMAM1989}, approximate inertial manifolds, which attract all orbits within a thin neighborhood of the global attractor, have been constructed in two spatial dimensions \cite{foias1988modelling}.
%The physical intuition behind the collapse to the approximate inertial manifold (hereafter referred to as the invariant manifold) is that small scales are rapidly damped due to viscosity. 
The inertial manifold encapsulates the essential features of the system, with an intrinsic dimension $d_{\mathcal{M}}$, typically much smaller than the dimension $d_{\omega}$ of the state space of the discretized PDE. 

%Simple dimensional arguments are consistent with rigorous upper bounds ( Clark 2020)
While the existence of an inertial manifold is not proven, the global attractor $\mathcal A_G \subset \mathcal M$ exists for the forced 2D Navier-Stokes equations \cite{temam1997infinite}, and its dimension has been bounded from above \cite{CONSTANTIN1988,DOERING1991} and below \cite{Liu1993} (the latter in monochromatic Kolmogorov flow).
% A rigorous upper bound for the dimension of the global attractor} has been derived \citep{CONSTANTIN1988}, and re-derived \cite{DOERING1991}, for the forced, two-dimensional Navier-Stokes equations on a doubly periodic, square domain.
The bounds are asymptotically tight (up to a logarithmic factor): The upper bound scales with a non-dimensional Grashof number 
$G := (\int_A|\mathbf f|^2 dA)^{1/2}\, L^2 / \nu^2$ % putting integral in because the norm here is dimensional 
% $G := \|\mathbf f\|_2 L^3 / \nu^2$ 
as $G^{2/3}(1 + \log G)^{1/3}$ \cite{CONSTANTIN1988,DOERING1991}, which depends on the forcing profile $\mathbf f$, box length $L$ and viscosity $\nu$, while the lower bound scales like $G^{2/3}$ \cite{Liu1993} (this computation is based on the dimension of the unstable manifold of the laminar fixed point).
%$G = \frac{|f|}{\nu^2 \lambda_1} = 4\pi^2 Re^2 (for KF non-dimensionalization) or just Re^2 (not fully sure)$
%Equivalently, this upper bound can be expressed in terms of the largest length scale in the domain $L$ and a small scale $l_d$, set by the bottom of the inertial range below which viscosity effects dominate \cite{Frisch_1995}.
Alternatively, a heuristic argument for the dimension of the inertial manifold has been made by determining the total number of excited modes in the inertial range of the (direct) turbulent cascade.
%This estimate then scales like $N \sim (L/l_d)^2$ in two-dimensions, where $L$ is the largest length scale in the domain and $l_d = \left( \nu^3 /\eta \right)^{1/6}$ is the dissipative length scale, where $\nu$ is the kinematic viscosity and $\eta = \nu \langle \nabla^2 u \rangle^2$ is the enstrophy dissipation rate. 
This estimate scales like $N \sim (L/l_d)^2 \propto G^{2/3}$ in two-dimensions, where $L$ is the largest length scale in the domain and $l_d$ is the dissipative (Kraichnan) length scale \cite{kraichnan1967}. 
This heuristic estimate was extended to $G^{2/3}(\log G)^{1/3}$ by incorporating a logarithmic correction to the energy spectrum to ensure a constant enstrophy flux in the direct cascade \cite{Kraichnan_1971}, and was shown to be consistent with the rigorous upper bound \cite{ohkitani1989}. 
%Bounds for square domain (Liu lower bound which is consistent with upper bound in Constantin, up to logarithmic factor). 
% don't think this lower bound is rigorous -- it is derived by estimating the number of unstable directions around this steady state.
%A lower bound of $G^{2/3}$, consistent with the upper bound up to a logarithmic factor, has also been derived for two-dimensional turbulence on a doubly periodic square domain, subject to a special class of forcing \cite{Liu1993}. 
%Bounds are for box with small aspect ratio ( Babin and Vishik lower bound, Ziane upper bound -- full agreement in upper and lower bound  )
%Upper \cite{ZIANE19971} and lower \cite{Babin1983} bounds with an equivalent scaling law have also been derived for two-dimensional turbulence in highly elongated, periodic domains.

% Robust estimates of $d_{\mathcal{M}}$ are essential for reduced order models of turbulent flows.  
Our focus in this work is on generating low-dimensional embeddings for the chaotic attractor $\mathcal A \subset \mathcal A_G$ in two-dimensional Kolmogorov flow.
\AC{$\mathcal A$ is defined as an invariant set which attracts an open set of initial conditions and is minimal -- no proper subset of $\mathcal A$ exists which satisfies the first two conditions.}
\AC{The chaotic attractor presumably requires far fewer degrees of freedom than $\mathcal A_G$, which must contain all exact coherent structures and their unstable manifolds.}
%which describes the long-time evolution of turbulence and which presumably requires far fewer degrees of freedom than $\mathcal A_G$.
To estimate the scaling $d_{\mathcal A}:=\dim(\mathcal A)$ with Reynolds number, $Re$, we construct highly-accurate autoencoder neural networks in an effort to find robust `latent' coordinate systems. 
Our objectives are twofold: (1) to estimate the $Re$-dependence of $d_{\mathcal A}$ and (2) to use the latent coordinate system to map typical dynamical pathways through different vortical `events' and to understand the coverage (or lack of coverage) of these dynamics by large libraries of recently computed unstable periodic orbits \cite{Page2024,page2022recurrent}.
% This work leverages an implicit rank minimization autoencoder framework \citep{jing2020,Zeng_2024} to estimate the scaling of an upper bound for $d_{\mathcal{M}}$ with Reynolds number ($Re$) in Kolmogorov flow \cite{Chandler2013}. 

% 2) data driven techniques applied to Kolmogorov flow

%Recent efforts to characterize these invariant manifolds of turbulent flows have leveraged ideas from dynamical systems and data-driven approaches.
%These efforts consider turbulent flows as one-dimensional curves on the invariant manifold, transiently approaching the neighbourhoods of exact solutions, such as equilibria, travelling waves (TWs) or unstable periodic orbits (UPOs) \cite{Hopf1948, Kawahara2012, Graham2021}. 
%Temporal averages of turbulent statistics in two-dimensional Kolmogorov flow have been reconstructed by weighted sums of the statistics of these exact solutions \cite{Chandler2013, lucas2015, page2022recurrent,Redfern2024}.

%Numerical computation results (Olson , Grappin x2 , Clark ). 
%[Andrew: could you verify some of the statements below? Just double check which cascade (forward or inverse) the authors were referring to in the varios cases. ]}
Numerical computations for the chaotic attractor dimension (by computing Lyapunov exponents and invoking the Kaplan-Yorke conjecture \cite{kaplan1979}) give some indications as to a scaling with $Re$. 
The Lyapunov dimension of the attractor of Kolmogorov flow was reported to depend strongly on the wavenumber $n$ chosen for the monochromatic forcing profile, and only weakly on $Re \propto G^{1/2}$ \cite{grappin1987}.  
%indicating that few wavenumbers larger than the forcing wavenumber $k_f$ are turbulent \cite{grappin1987}. 
% linear scaling in the case of artificial large scale damping to set up the energy inertial range -- I think best to avoid discussing this. 
% in hindsight, this sounds a lot like Masa's findings...
A follow-on study of Kolmogorov flow found that the Lyapunov dimension is bounded above by the number of degrees of freedom contained in scales larger than the forcing scale, $N_{LS}\propto n^2$
% this study also considered artificial large and small scale damping -- found a linear dependence of dimension on number of modes in inverse energy inertial range.
% their Re is the same as the Platt Re i.e. Re_{theirs} = 8 Re_{ours} -- easy to derive, just remember that Grappin define streamfunction forcing so our \chi = their f * n.
\cite{Grappin_1991}, though this study considered relatively weak turbulence with $16.4\leq Re \leq 46.2$ in our units.
\AC{A scaling of the Lyapunov dimension $\sim Re^{0.78}(n / k_0)^2$ was also computed numerically for forced two-dimensional turbulence with large scale damping, where $k_0 = 2\pi / L$ \cite{clark2020}, also demonstrating a linear dependence on $N_{LS}$. 
}
%[I don't quite understand this last point, and also what is the inertial range referred to here?]} 

% 4) Previous numeric approaches to estimate dimension of pdes
Related to these direct numerical estimates, there has been a recent focus on using data-driven approaches to estimate the dimensionality of the inertial manifold of simpler chaotic systems.
Linear approaches, such as principal component analysis (PCA) and its extensions (e.g., Sparse PCA \cite{Zou2006}, Bayesian PCA \cite{bishop1998}), tend to overestimate the dimension as they are limited to linear subspaces. 
%They identify a linear subspace where the projection error is minimized. 
%These methods are computationally efficient and yield explicit mappings to reduced-dimensional spaces. 
%However, because they are limited to linear subspaces, they tend to overestimate the intrinsic dimension $d_{\mathcal{M}}$, as representing data on a curved manifold generally requires at least $d_{\mathcal{M}} + 1$ dimensions. 
Nonlinear techniques (e.g. deep autoencoders) address this limitation to more faithfully learn the structure of the inertial manifold.
By examining how the model reconstruction error of the data changes with the dimensionality of the embedding space, autoencoder networks have been used to identify the inertial manifold dimension of the Kuramoto–Sivashinsky (KS) equation \cite{Linot2020, linot2022, vlachas2022}. 
%Nonlinear PCA has shown promise in fluid dynamics, with Milano and Koumoutsakos \cite{MILANO2002} demonstrating improvements in reconstruction error for the randomly forced Burgers equation and turbulent channel flow compared to PCA. 
% Despite these successes, visualizing nonlinear PCA modes remains challenging due to the nonlinear coupling of temporal and spatial features. 
% Recent efforts have proposed solutions, such as the additive output layer by Fukami et al. [8], which facilitates the interpretation of spatial structures, and the latent Fourier analysis by Page et al. [19], which reveals patterns by filtering latent wavenumbers. 
However, this approach requires a manual sweep over multiple latent dimensions and there is no apparent drop-off in reconstruction error for more complex systems and higher spatial dimensions \cite{dejesus2023_kf, Linot_Graham_2023}. 
% IRMAE networks
A recent study has proposed a more systematic approach to determine the dimensionality of the inertial manifold of the KS equation, using the so-called Implicit Rank Minimizing Autoencoder with Weight Decay (IRMAE-WD) \cite{Zeng_2024}.  
The design of this architecture exploits the phenomenon that gradient-based optimization of deep \textit{linear} neural networks leads to low-rank solutions (implicit rank minimization) \cite{jing2020}, while the weight decay refers to an $L_2$ regularization on the weights of the network to further drive down the rank of the learned embedding. 
The same network was applied to both two-dimensional Kolmogorov flow with a forcing wavenumber of $n = 2$ at low $Re$ \cite{dejesus2023}, and two-dimensional Rayleigh-Bénard flow \citep{Vinograd_Clark_2025} to compute estimates for the dimensionality of these flows.

% 5) Symmetry considerations
% When determining the dimensionality of invariant manifolds of chaotic systems, it is important to respect the symmetries of the governing equations. 
Many canonical fluid flows exhibit continuous and/or discrete symmetries in the governing equations, which must be accounted for in the training of a nonlinear neural network.
% Otherwise, a data-driven reduced order model will likely not learn an optimal representation of symmetric copies of solutions to the governing equations.
The simplest way to encourage the model to learn the symmetries of the equations is via data augmentation, where each symmetry is randomly applied to the training data after each epoch \cite{Page2024, krizhevsky2012} to ensure a robust sampling of the symmetry groups.
However, this approach does not guarantee equivariance of the model with respect to the symmetries. 
% discrete and continuous symmetries exist
In this work, to explicitly reduce a continuous or discrete symmetry, each snapshot is mapped a symmetry-reduced subspace of the full state space. %, such that the group orbit of the symmetry intersects just once with this submanifold.
%To explicitly reduce a discrete symmetry with $n$ symmetric copies of each snapshot, we partition the state space into $n$ subdomains.
This pre-processing of the data restricts the model to learning a representation within this subspace \cite{zeng2021, dejesus2023}.
The first Fourier mode method of slices can be used to explicitly reduce continuous symmetries of dynamical systems \cite{budanur2015}. 
% dejesus for discrete symms
%Discrete symmetries require a more ad-hoc formulation of the fundamental symmetry domain.
% For discrete symmetries,} A symmetry reduction scheme for the discrete reflection symmetry in the Kuramoto-Sivashinsky equation was formulated by constructing polynomial invariants from Fourier modes \cite{Budanur_2016}. 
For discrete symmetries, we adapt a reduction technique that defines an indicator function based on Fourier modes to separate the state space of the governing equation into symmetric charts, and maps all solutions into one of these charts, defining a `fundamental chart' \cite[this was applied to the KS system in][]{zeng2021}. 
This method of `symmetry charting' has been applied to Kolmogorov flow with forcing wavenumber $n = 2$ \cite{dejesus2023}, where an indicator function was defined on specific Fourier modes to define eight symmetry charts representing the eight-fold discrete symmetries. %, one of which is defined as a fundamental chart. 
Other work has explicitly accommodated various discrete symmetries via the use of many identical (`Siamese') networks to identify a fundamental domain \cite{kneer2021}.

Other studies have also considered the problem of simulating dynamics in the latent space of an autoencoder -- motivated by the extreme computational requirements of direct numerical simulation of turbulence at high $Re$.
% As turbulent flows are characterized by multiple temporal and spatial scales, direct numerical simulations of the Navier-Stokes equations are typically computationally prohibitive due to resolution requirements.
% A full characterization of the invariant manifold, such that both an invertible mapping from phase space to the manifold and the time-forward map on the manifold are learned to machine precision, would greatly reduce this computational cost. 
% Although the robust estimation of the dimensionality of the invariant manifold of two-dimensional turbulence is still an outstanding challenge, a flurry of reduced order models have been constructed in recent years.
%An estimate of the dimension of the invariant manifold would be beneficial in determining the minimum number of degrees of freedom for these reduced order models.
Kolmogorov flow has been a popular testing ground for these efforts to date. 
For example, convolutional autoencoders have yielded a latent space representation which revealed a clear separation between the high dissipation bursting dynamics and the low dissipation quiescent dynamics, at moderate Reynolds number $Re$ \cite{Page2021, Page2024}. 
Autoencoders have also been combined with echo state networks as a reduced order model to accurately predict the flow at different $Re$ \cite{Racca_Doan_Magri_2023}.
%The latent space learned by autoencoders is a curved manifold with curvilinear coordinates, which can be investigated with tools from differential geometry \cite{magri2022}.
%The numerical computation of geodesics \cite{kelshaw2024} allows for the generalization of proper orthogonal decomposition to this low-dimensional latent manifold, yielding a dominant mode which represents physical structures typical in two-dimensional Kolmogorov flow.
An imperfect reduced order model for Kolmogorov flow has been complemented with a data-driven correction term to accurately predict extreme high dissipation events \cite{wan2018}.

% will investigate the importance of factoring out the symmetries in the work
In this work, we will combine convolutional `dense blocks' \cite{Page2024}, the IRMAE-WD network approach \cite{jing2020,zeng2021} and a full symmetry reduction step via symmetry charting, to estimate the scaling with $Re$ of an upper bound to the chaotic attractor dimension of the $n = 4$ Kolmogorov flow.
%We investigate how this estimate scales with $Re$, and make comparisons with the previous analytic scaling bounds and numerical estimates. 
The combination of the IRMAE model with a full symmetry reduction yields a very rich and interpretable learned latent space, which reveals new physical insight to the dynamics of the flow. 
In section \ref{sec:flow_conf}, we formulate the equations of motion of two-dimensional Kolmogorov flow and describe the data generation and symmetry reduction.
In section \ref{sec:network}, we describe the autoencoder architecture.
In section \ref{sec:dim}, we report the estimated scaling of the chaotic attractor dimension with $Re$.
In section \ref{sec:latent_structure}, we investigate the structure and modal decomposition of the learned embedding spaces of the models. 
%We also investigate the addition of a dynamical term in the loss function of the trained IRMAE-WD model, in which the time evolution of the true Kolmogorov solution is faithfully captured by the time-forward map of the reconstructed initial condition output by the autoencoder. 

\section{Flow configuration}
\label{sec:flow_conf}

\subsection{Numerical simulations}
\label{sec:flow}

%formulate the equations of kf
We consider two-dimensional turbulence on a doubly periodic domain, driven by a monochromatic body force in the streamwise direction (Kolmogorov flow). 
% This well-studied flow configuration is known as Kolmogorov flow \citep{Chandler2013}.
The out-of-plane vorticity $\omega = \partial_x v - \partial_y u$, where the velocity $\bm{u} = (u,v)$, evolves according to
\begin{equation}
    \partial_t \omega + \bm{u} \cdot \boldsymbol\nabla{\omega} = \frac{1}{\Rey} \nabla^2 \omega - n \cos {ny}.
    \label{eq:kf_eq}
\end{equation}
In this non-dimensionalization, a length scale is chosen as the inverse of the fundamental wavenumber of the box $1/k^* = L^* / 2\pi$, where $L^* = L^*_x = L^*_y$ (i.e.\ equal aspect ratio -- the asterisk denotes dimensional units here).
A time scale is chosen as $1/\sqrt{k^*\chi^*}$, where $\chi^*$ is the amplitude of the forcing in the momentum equation. 
These length and time scales define the Reynolds number for this configuration $\Rey \coloneq \sqrt{\chi^*/k^{*3}}/\nu$, where $\nu$ is the kinematic viscosity.
Throughout this work, the forcing wavenumber is set to $n = 4$, as has been common in previous studies \citep{Chandler2013, Page2021, page2022recurrent}.
% symmetries of kf
Equation (\ref{eq:kf_eq}) with periodic boundary conditions and $n=4$ forcing waves is equivariant under the action of three symmetries:
\begin{enumerate}
    \item continuous shifts in the streamwise ($x$) direction, $\mathscr{T}_s : \omega(x,y) \to \omega(x+s, y)$,
    \item discrete shift-reflects by a half-wavelength in $y$, $\mathscr{S}: \omega(x,y) \to -\omega(-x, y+\pi/4)$ and
    \item discrete rotations by $\pi$, $\mathscr R : \omega(x,y) \to \omega(-x,-y)$, 
\end{enumerate}
which we account for explicitly when estimating the dimension of the turbulent attractor.

Key measures of the flow used throughout this study are
%the total kinetic energy
%\begin{equation}
%    E(t) \coloneq \frac{1}{2} \langle \bm{u}^2 \rangle_V,
%\end{equation}
the total dissipation rate,
\begin{equation}
    D \coloneq \frac{1}{\Rey} \langle | \nabla\bm{u} | ^2 \rangle_V, %= \frac{1}{\Rey} \langle \omega^2 \rangle_V,
\end{equation}
and the total production rate,
\begin{equation}
    I \coloneq \langle u \sin(ny) \rangle_V,
\end{equation}
where the average over the volume $V=[0,2\pi]^2$ is defined as
\begin{equation}
    \langle \bullet \rangle_V \coloneq \frac{1}{(2\pi)^2} \iint_V  \, \bullet \, d^2\bm{x}.
    \label{eq:spatial_av}
\end{equation}

We consider this flow at a number of $Re \in \{$25, 30, 35, 40, 60, 80, 100, 200, 400$\}$. 
Kolmogorov flow approaches an asymptotic regime \citep{Chandler2013, cleary2025} with $D \sim Re^{-1/2}$ beyond $Re \approx 50$. 
A variety of simpler states have been reported at lower $Re$ \citep{Chandler2013}.
The majority of $Re$ values considered are sampled from the asymptotic chaotic regime.
We will devote particular attention to the well-studied values $Re = 40$  and $Re = 100$. %here, which is considered a transitively turbulent flow.
% Some $Re$ values between the meta-stability regime and the asymptotic regime are also considered.

% time integrator
We solve equation (\ref{eq:kf_eq}) using the spectral version of the \texttt{JAX-CFD} solver \citep{Kochkov2021, LCspectral}.
The velocity field is computed at each time-step via solution of the Poisson equation $\nabla^2 \psi = -\omega$, where the streamfunction $\psi$ is related to the induced velocity components via $u = \partial_y \psi, v = -\partial_x \psi$.
%\texttt{JAX-CFD} is a fully-differentiable solver, which enables one to compute gradients of loss functions involving the time-forward map $f^t(\omega)$ of equation (\ref{eq:kf_eq}) to machine precision, via automatic differentiation.

\subsection{Datasets}

% datasets and resampling
For $Re \leq 200$, training data was generated
% The training dataset was generated at the considered values of $\Rey \leq 200$ by
from 400 trajectories, each of length $10^3$ advective time units, with an initial transient period of 50 advective time units discarded.
At $Re = 400$, we instead computed 4000 trajectories each of length $10^2$ (to enable parallelization over trajectories and avoid excessive compute time at high resolution).
Data was sampled every advective time unit ($\Delta t^* = 1 / \sqrt{k^*\chi^*}$ in dimensional units), resulting in a dataset of $4\times 10^5$ individual snapshots for each $Re$ value considered.
A spatial resolution of $128^2$ was sufficient for $Re < 200$, while trajectories at $Re = 200$ and $400$ were computed with $256^2$ and $512^2$ gridpoints respectively, before being downsampled to $128^2$ for input to the autoencoders.
We reserved 10\% of the training data as a `validation' dataset, which is not used to update the model weights, but instead is used to track convergence metrics throughout training.
Test datasets of $5\times 10^4$ snapshots for each $Re$ were generated in the same manner.

\subsubsection{Resampling}

\begin{figure}%[!ht]
    \centering
    \includegraphics[width=\linewidth]{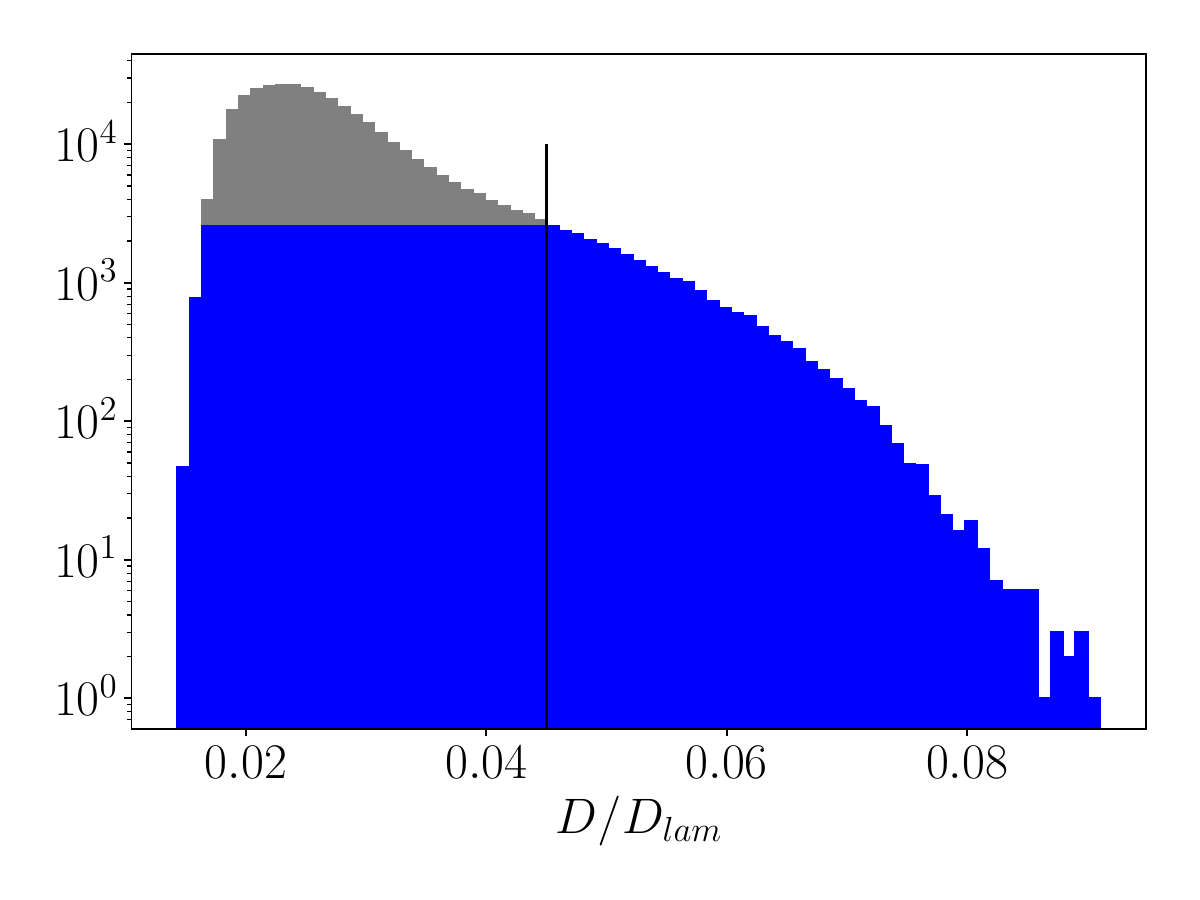}
    \caption{The distributions of dissipation rate $D$ at $Re = 100$, normalized by the laminar dissipation rate $D_{lam} = Re / (2n^2)$, is shown for the full training dataset (gray) and the resampled training dataset (blue).
    % not sure we need the following
    % Neither distribution has been normalized to integrate to 1, allowing for clear comparison. [[-- not sure if we need this statement?]]}
    %The new distribution has not been normalized to allow for comparison to the unmodified data.} 
    }
    \label{fig:resampling_strategy}
\end{figure}

Training models on data sampled from the turbulent attractor, as described above, lead to low reconstruction errors for low-dissipation-rate snapshots, but high errors for high-dissipation-rate snapshots as observed in previous studies \citep{Page2021,Page2024}. 
This is because (1) low-dissipation snapshots are less spatially complex and (2) high-dissipation events are rare events in Kolmogorov flow \cite{Chandler2013, farazmand2017} -- the dataset is dominated by low-dissipation events (see figure \ref{fig:resampling_strategy}). 
One approach that has previously been used to alleviate this effect is the introduction of a loss term involving the square of the vorticity field \cite{Page2024} to add additional penalization to higher dissipation events.  
Here, we instead introduce a resampling strategy to alter the distribution of training data over dissipation and artificially increase the frequency of high dissipation events.  
To do this, the distribution of dissipation rate is first partitioned into `low' and `high' dissipation events, defined by a threshold value $D_c$. 
For example, at $\Rey = 100$, this value is set to $D_c = 0.045 D_{lam}$ (the laminar dissipation rate is $D_{lam} = \Rey / 2n^2$).
% The idea is to resample the data so that the high dissipation events are more populous throughout the dataset.
All data above the threshold value is sampled from the original training dataset.
The dataset is then binned according to dissipation rate, with bin widths of $\Delta D = 0.001 D_{lam}$ for $Re < 200$, $\Delta D = 0.0005 D_{lam}$ for $Re = 200$ and $\Delta D = 0.0001 D_{lam}$ for $Re = 400$. 
To smoothly reduce the number of low dissipation events in the dataset, a maximum of $n_l$ snapshots are then randomly sampled from each bin below the threshold $D_c$, where $n_l$ is set by the number of samples in the first `high' dissipation bin $[D_c, D_c + \Delta D]$. 
%$N/2$ snapshots are sampled from the full dataset such that the resampled distribution of the $N$ snapshots is approximately uniformly distributed below this threshold. 
The modified distribution created by this resampling strategy is shown in figure \ref{fig:resampling_strategy} for $Re = 100$.

\AC{The parameter $D_c$ \JP{must balance two key considerations}: that (1) the network is exposed to a relatively high number of high-dissipation snapshots and that (2) a sufficiently large training dataset \JP{is retained after resampling from the} $4\times10^{5}$ \JP{original} snapshots at each $Re$. 
These are competing conditions as the size of the resampled dataset can be increased by lowering $D_c$, but \JP{with a reduced} weighting given to the high dissipation snapshots.}
The thresholds, in order of increasing $Re$, are set at $D_c/D_{lam} \in \{$0.16, 0.1415, 0.1255, 0.15, 0.125, 0.0725, 0.045, 0.0125, 0.0045$\}$.
As a result of the resampling, the size of the training dataset is reduced, with a total number of snapshots snapshots $N_S \approx (20, 40, 50, 60, 75, 90, 100, 125, 130)\times 10^3$ (ordered in increasing $Re$).

\subsubsection{Symmetry reduction}
\label{sec:sym_red}

As described in §\ref{sec:flow}, Kolmogorov flow is equivariant under continuous shifts in the streamwise direction, under discrete shift-reflects by a half-wavelength in $y$, and under rotations by $\pi$.
We account for all symmetries explicitly in a pre-processing step before training the neural networks.
The continuous translational symmetry is reduced by the first Fourier mode method of slicing \cite{budanur2015}. 
In this method a phase $\phi$ is computed such that the $(k_x,k_y) = (1,0)$ Fourier mode of the translated state $\mathscr T_{-\phi}\omega$ is purely real.
% This method considers the spatial phase $\phi$ of the $(k_x, k_y) = (1, 0)$ Fourier mode, and translates $\omega$ by $-\phi$ so that this mode of the translated state $\mathscr T_{-\phi}\omega$ is purely real.

For the discrete symmetries, we adapt the symmetry charting approach introduced in \cite{dejesus2023} for $n=2$ Kolmogorov flow.
% We present here the symmetry charting approach used to reduce the discrete symmetries of $n = 4$ Kolmogorov flow, inspired by the method presented for $n = 2$ Kolmogorov flow in \cite{dejesus2023}. 
This approach partitions the full state space into symmetrically related charts, and assigns each chart a unique indicator $\mathscr{I}$. 
% By carefully tabulating combinations of symmetry operations to map states between charts, it is possible to identify a fundamental chart to which all states should be mapped. 
% I don't think there is a "fundamental chart" where things "should" be mapped, I think we identify a chart and map to it, right? 
We select a single chart to which we map all states by combinations of $\mathscr{S}$ and $\mathscr{R}$, which is done by looking at specific terms in the Fourier transform of the vorticity field,
% We define the indicators on specific Fourier modes $\hat{\omega}(k_x, k_y)$, defined by the Fourier transform
$$\omega(x,y) = \sum_{k_x} \sum_{k_y} \hat{\omega}(k_x, k_y) e^{i(k_x x + k_y y)}.$$
The actions of the three symmetry operations on the Fourier modes are % given by the operators
\begin{align}
    \mathscr{T}_s : \hat{\omega}(k_x, k_y) &\to \hat{\omega}(k_x, k_y)e^{-ik_x s}, \label{eq:t_f} \\
    \mathscr{S}: \hat{\omega}(k_x, k_y) &\to -\hat{\omega}(-k_x, k_y)e^{ik_y\pi / n}, \label{eq:s_f} \\
    \mathscr R : \hat{\omega}(k_x, k_y) &\to \overline{\hat{\omega}}(k_x, k_y), \label{eq:r_f}
\end{align}
where $\overline{\bullet}$ denotes complex conjugation. 
%A Fourier mode will be chosen to reduce each of the three symmetries of Kolmogorov flow.
It is clear from (\ref{eq:t_f}) that the action of $\mathscr T_s$ only changes modes with $k_x \neq 0$.
However, $\mathscr R$ changes all modes, while $\mathscr S$ changes all modes except for modes $(0, (2l+ 1)n)$, where $l \in \mathbb{Z}$. 
As $\mathscr T_s$ does not affect modes with $k_x = 0$, we will use specific $k_x = 0$ modes to specify the indicators of the discrete symmetries, and reduce the discrete symmetries before the translational symmetry.
Otherwise, if the discrete symmetries are reduced after the translational symmetry, the (1,0) mode is no longer guaranteed to be purely real.

\begin{figure}
    \centering
    \includegraphics[width=\linewidth]{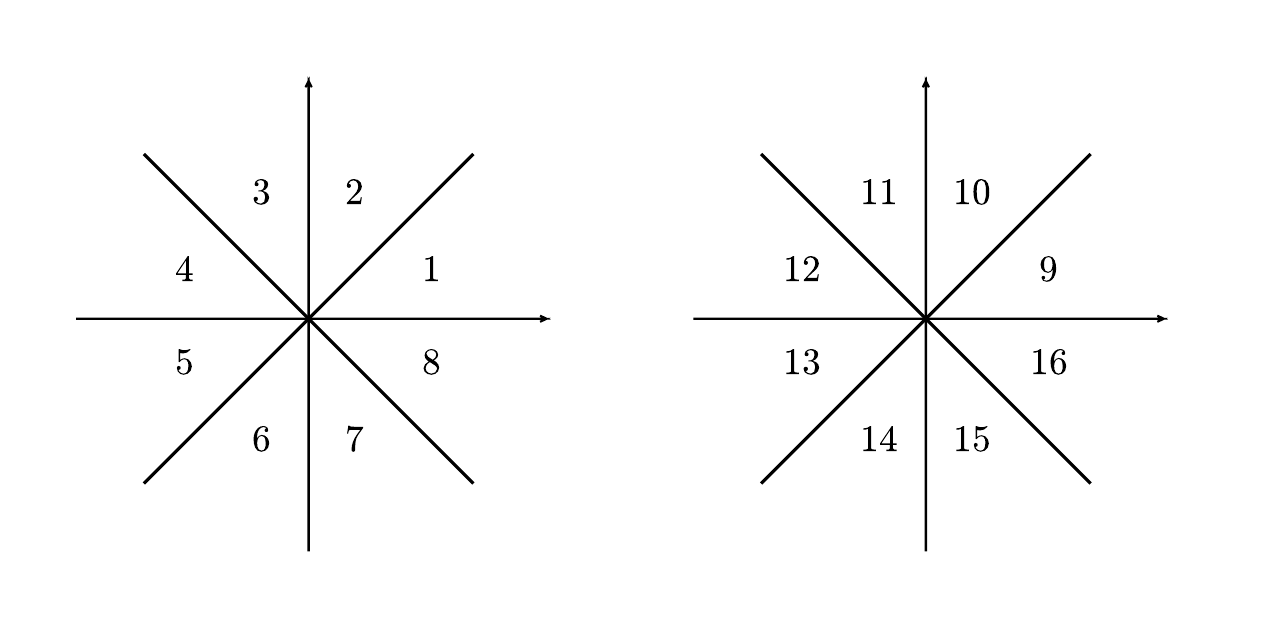}
    \caption{16 symmetry charts of $n = 4$ Kolmogorov flow, for each of the 16 discrete symmetric copies of a state. The $x$- and $y$-axes denote $\hat{\omega}_R(0,1)$ and $\hat{\omega}_I(0,1)$ respectively. (Left) $\hat{\omega}_I(0,4) > 0$ and (right) $\hat{\omega}_I(0,4) < 0$.}
    \label{fig:indicators}
\end{figure}

We consider the $(0,1)$ mode to diagnose the shift-reflect symmetry, splitting the complex plane spanned by the real $\hat{\omega}_R(0,1)$ and imaginary $\hat{\omega}_I(0,1)$ components of this mode into octants, representing the 8 shift-reflected copies generated by (\ref{eq:s_f}). 
We consider the $(0,4)$ mode for the rotational symmetry, splitting this complex plane into upper and lower halves, i.e.\ defined by the sign of $\hat{\omega}_I(0,4)$. 
Together, this decomposition results in 16 indicators as indicated in figure \ref{fig:indicators}, representing the 16 symmetric copies generated by both (\ref{eq:s_f}) and (\ref{eq:r_f}).
To reduce the discrete symmetries of a state, we select the `fundamental' domain as $\mathscr{I} = 1$, \AC{which is defined by $0< \arg \hat{\omega}(0,1) < \pi/4$ and $\hat{\omega}_I(0,4) > 0$}.

The required combination of symmetry operations to map a state from any $\mathscr{I}$ to $\mathscr{I} = 1$ is tabulated in table \ref{tab:maps}.
\AC{In practice, the symmetry reduction of a snapshot proceeds as follows: (i) identify the domain $\mathscr I$ of the snapshot from the $(0,1)$ and $(0,4)$ modes as in figure \ref{fig:indicators}, (ii) apply the appropriate symmetry operations from table \ref{tab:maps} to the snapshot, (iii) reduce the translational symmetry with the first Fourier mode method of slicing.}
Generalization of this approach to any number of forcing waves $n$ is relatively straightforward -- it requires the partitioning of the complex plane spanned by $\hat{\omega}_R(0,1)$ and $\hat{\omega}_I(0,1)$ into $n$ equally-sized sectors, along with consideration of the sign of $\hat{\omega}_I(0,n)$ to diagnose rotation. %\showthe\font

\begin{table}%[!hb]
    \begin{minipage}{0.5\linewidth}
        \centering
        \begin{tabular}{c|c}
        $\mathscr{I}$ & Action  \\ \hline
        1 & $\mathbb{I}$\\
        2 & $\mathscr{S}^3$\\
        3 & $\mathscr{S}^6$\\
        4 & $\mathscr{S}$\\
        5 & $\mathscr{S}^4$\\
        6 & $\mathscr{S}^7$\\
        7 & $\mathscr{S}^2$\\
        8 & $\mathscr{S}^5$\\
    \end{tabular}
    \end{minipage}%
    \begin{minipage}{0.5\linewidth}
        \centering
        \begin{tabular}{c|c}
        $\mathscr{I}$ & Action  \\ \hline
        9 & $\mathscr{S}^5 \mathscr{R}$\\
        10 & $\mathscr{S}^2 \mathscr{R}$\\
        11 & $\mathscr{S}^7 \mathscr{R}$\\
        12 & $\mathscr{S}^4 \mathscr{R}$\\
        13 & $\mathscr{S} \mathscr{R}$\\
        14 & $\mathscr{S}^6 \mathscr{R}$\\
        15 & $\mathscr{S}^3 \mathscr{R}$\\
        16 & $\mathscr{R}$\\
    \end{tabular}
    \end{minipage}
    \centering
    \caption{Symmetry operations to map all 16 symmetry charts of $n = 4$ Kolmogorov flow to the fundamental chart $\mathscr{I} = 1$}
    \label{tab:maps}
\end{table}

\section{Network architecture}
\label{sec:network}

\begin{figure*}
    \centering
    \includegraphics[width=\linewidth, trim={0 9.2cm 0 8.2cm}, clip]{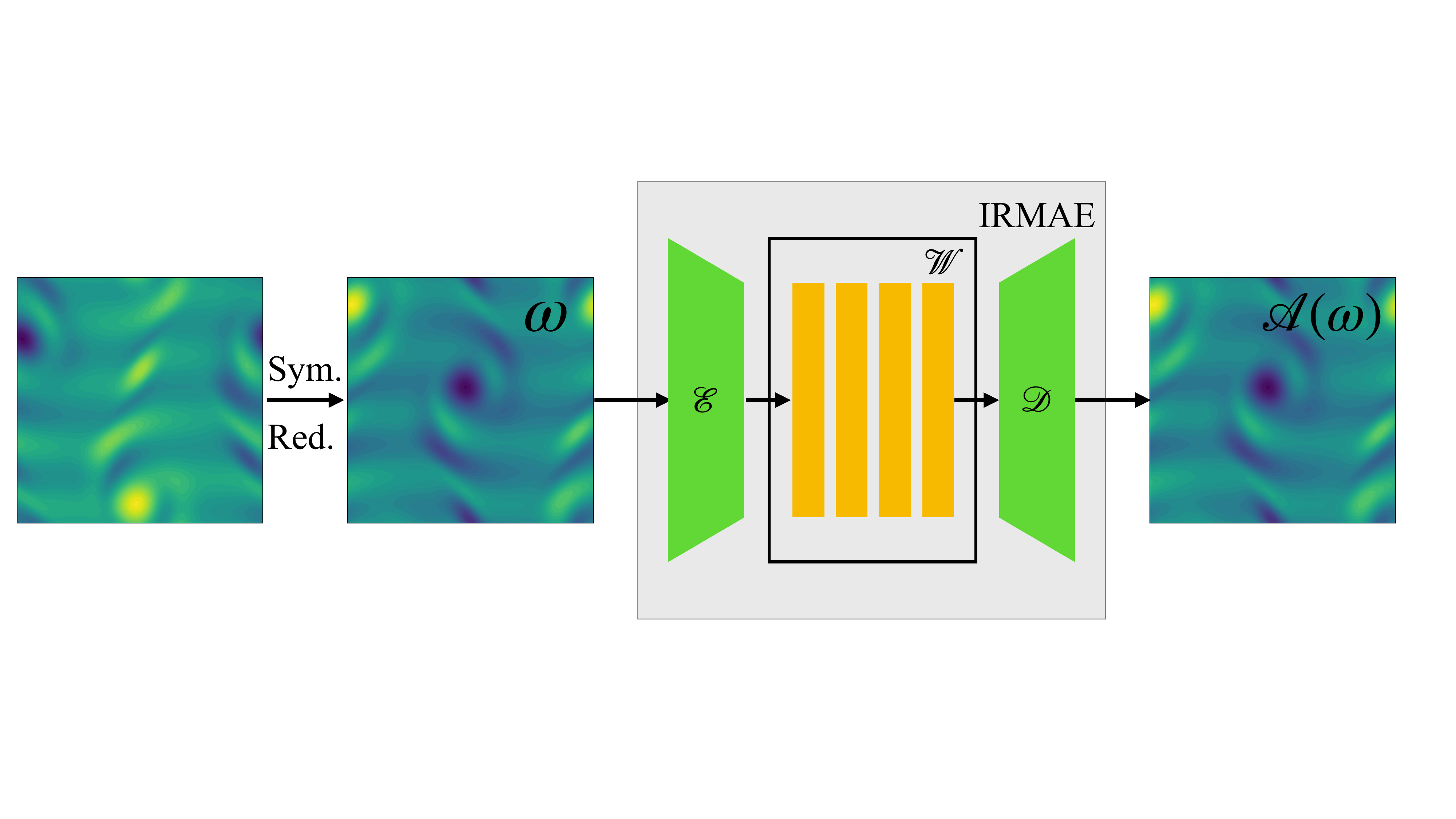}
    \caption{\AC{Schematic of the data/neural network pipeline used in this study. Symmetries are first reduced for each snapshot $\omega$, before being passed to the IRMAE network. The implicit rank minimizing layers (orange) are linear fully connected layers, while the encoder $\mathcal{E}$ and decoder $\mathcal{D}$ are fully described in Appendix \ref{app:arch_details}.}}
    \label{fig:irmae_schem}
\end{figure*}

We construct low-\AC{dimensional} representations of the turbulent attractor in Kolmogorov flow by combining the DenseNet convolutional autoencoder architecture from a previous study \cite{Page2024} (hereafter PHBK) with the implicit rank minimization and weight decay introduced by the IRMAE-WD architecture \cite{Zeng_2024, dejesus2023}.
Each deep convolutional autoencoder $\mathscr{A}$ seeks to reconstruct its input such that 
\begin{equation}
    \mathscr{A}(\omega) \equiv [\mathscr{D} \circ \mathscr{W} \circ \mathscr{E}](\omega) \approx \omega,
\end{equation}
where the encoder $\mathscr{E} : \mathbb{R}^{N_x\times N_y} \to \mathbb{R}^{d_z}$ maps the input vorticity snapshot to a low-dimensional representation, $\mathscr{W} : \mathbb{R}^{d_z} \to \mathbb{R}^{d_z} $ represents a series of $n$ fully-connected, equally-sized linear layers (pure matrix multiplication) within the embedding space such that $\mathscr{W} = \mathbf W_1 \mathbf  W_2 \cdots \mathbf W_n$ and $\mathbf W_i \in \mathbb{R}^{d_z\times d_z}$, and the decoder $\mathscr{D} : \mathbb{R}^{d_z} \to \mathbb{R}^{N_x\times N_y}$ maps the embedding back to a vorticity snapshot. 
\AC{A high-level schematic of this architecture and the symmetry reduction pre-processing step is presented in figure \ref{fig:irmae_schem}.}

The structures of the encoder $\mathscr{E}$ and decoder $\mathscr{D}$ considered here are adapted from PHBK, with so-called `dense blocks' \cite{Huang2016, Huang2019} instead of single convolutional layers.
The dense blocks allow for much richer feature maps at each scale throughout the autoencoder.
Full details on the architecture used here are given in appendix \ref{app:arch_details} \AC{and the full source code is available at \url{https://github.com/computational-flow-science/IRMAE}.}
%All differences between the architecture used in this work and in PHBK are outlined here;
%The dense blocks in this work only consist of two convolutional layers, rather than three as in \cite{Page2024}, as the symmetry reduction preprocessing allows for a reduction in model complexity.
The key addition to the architecture is the insertion of fully-connected linear layers into the bottleneck of the autoencoder.
The implicit rank minimization of these linear layers drives the output of $\mathscr{W}(\mathscr{E}(\omega))$ to be of minimal rank \citep{jing2020}. 
% (1) Fewer max pooling operations are performed in the architecture considered here, such that the output of the encoder is an `image' of shape (16, 16, 4), which is then flattened to a vector of length $d_z = 1024$. 
% (2) The insertion of fully-connected linear layers into the bottleneck of the autoencoder considered here drives the output of $\mathscr{W}(\mathscr{E}(\omega))$ to be of minimal rank by implicit rank minimization. 
% This embedding is then reshaped back to shape (16, 16, 4), and fed to the decoder to generate the vorticity field reconstruction. 
% (3) The final convolutional layer with kernel size $(8,8)$ and tanh activation function in PHBK is replaced by a sequence of linear convolutional layers of decreasing kernel sizes from $(16,16)$ to $(2,2)$.
% The wide range of kernel sizes allows for a more faithful reproduction of the small spatial scales. 
%The input data does not need to be normalized by some constant $\omega_{\text{norm}}$ due to the linearly activated output, as was required by the final tanh activation in PHBK.
The densely connected linear layers to the autoencoder bottleneck do not increase the expressivity of the network, as these layers can be expressed as a multiplication by a square matrix. 
However, the number of parameters increases to $\sim 6.21 \times 10^{6}$ trainable parameters due to the insertion of these linear layers, nearly triple the $\sim 2.15 \times 10^6$ trainable parameters in the comparable models considered in PHBK.

\AC{Although more complex (generative) approaches have been shown to learn informative latent representations of chaotic dynamics \citep{SoleraRico2024, li2024}, the IRMAE network -- while fully deterministic -- has been shown to result in latent (or embedding) spaces with similar favorable properties.
For example, comparable performance to variational autoencoders has been demonstrated in tasks such as smoothly interpolating in the latent space and generating new samples from random noise \cite{jing2020}. 
Furthermore, while the latent spaces of variational autoencoders are typically full rank, the IRMAE networks benefit from approximately minimal rank latent spaces, which are key in estimating the dimensionality of the chaotic attractor.
}

% loss function + weight decay
The networks are trained to minimize the standard mean squared error loss function
\begin{equation}
    \mathscr{L} = \frac{1}{N_S} \sum_{j=1}^{N_S} \| \mathscr{A}(\omega_j) - \omega_j \|^2,
    \label{eq:loss}
\end{equation}
% old spectral loss function
% \begin{equation}
%     \mathscr{L} = \frac{1}{N} \sum_j \| \mathscr{A}(w_j) - w_j \|^2 + \alpha \mathscr{L}_{\text{spec}},
%     \label{eq:loss}
% \end{equation}
%where $\theta$ denotes the trainable parameters of the autoencoder, such that the final term implements the $L_2$ weight decay.
\AC{where each $\omega_j$ is a snapshot from the training dataset}, using an AdamW optimizer \cite{loshchilov2019} with an initial learning rate of $10^{-4}$ and the default weight decay coefficient of 0.004.
This optimizer decouples the weight decay from the adaptive gradient updates, as including an explicit $L_2$ regularization term on the trainable parameters in (\ref{eq:loss}) is not effective with Adam-style optimizers \cite{Kingma2015}, nor identical to weight decay \cite{loshchilov2019}.
The batch size is set to 8 for $Re \le 40$, 12 for $Re = 60$ and 16 for $Re \ge 80$.
Setting relatively small batch size increases the total training time, but also increases the total number of gradient-based optimization steps which we found to be effective in driving down the rank of the learned embedding. 
The networks are trained with the AdamW optimizer for at least 1000 epochs for each $Re$.
Inspired by previous studies implementing a two-optimizer training process \citep{WANG2024}, the best model to that point is then further tuned using the AdaGrad optimizer \citep{Duchi2011} for a further 1000 epochs.
\AC{A brief hyperparameter analysis over the initial learning rate and batch size is presented in appendix \ref{app:hyperparam}.}
\JP{The model size and associated training time (2 weeks of GPU time per model reported here) prohibited us from performing an extensive hyperparameter sweep over all $Re$, and better performance may well be possible with alternative settings.}

% other attempts to improve reconstruction loss
Alternative loss functions, for instance asserting that the time evolved reconstructed vorticity fields must match the time evolved ground truth \citep{Page_2025}, and spectral regularization terms which penalize poor energy spectrum reconstruction \citep{Durall2020}, were also considered here but did not result in any substantial improvements to the reconstruction error. 
We also attempted to fit a second-stage autoencoder network, sharing the same latent representation of the data, to the residual of the reconstructed vorticity fields, motivated by the architecture used to perform machine precision reconstruction of a single turbulent snapshot \citep{ng2024}.
However, the error did not appear to have any structure to learn and the second-stage autoencoder was unable to reduce the error any further.
% However, the residual fields were found to be mostly white noise, to which the second-stage autoencoder could not be fit. 

%
% spectral regularization discussion
%
%A spectral regularization term $\mathscr{L}_{\text{spec}}$ is added to the loss function with a weighting of $\alpha$, which was found to prevent overfitting and improve reconstruction errors in the test set. 
%The spectral regularization term used here is motivated by the spectral term added to the generator loss function in \cite{Durall2020}, which showed that a correct approximation of the frequency spectrum improves the training stability and quality of generative adversarial networks. 
%Letting $E_{w}(k)$ denote the enstrophy spectrum of the training dataset, the spectral regularization term considered here seeks to match the reconstructed enstrophy spectrum $E_{\mathscr{A}(w)}(k)$ as closely as possible with $E_{w}(k)$, such that
% \begin{equation}
%     \mathscr{L}_{\text{spec}} = \frac{1}{k_{max}} \int_0^{k_{max}} ( E_{\mathscr{A}(w)} - E_{w} )^2 dk.
%     \label{eq:loss_spec}
% \end{equation}
% reconstruction error and results -- compare resampling and regularization 

\section{Dimension estimation}
\label{sec:dim}

In this section, we first report the reconstruction error of our model across all $Re$ considered.
We then present the estimated scaling for (an upper bound on) the dimension of the chaotic attractor with $Re$. 
%The reconstruction error indicates a good approximation of the invariant manifold across all $Re$,  

\subsection{Reconstruction error}

\begin{figure}%[!ht]
    \includegraphics[width=\linewidth]{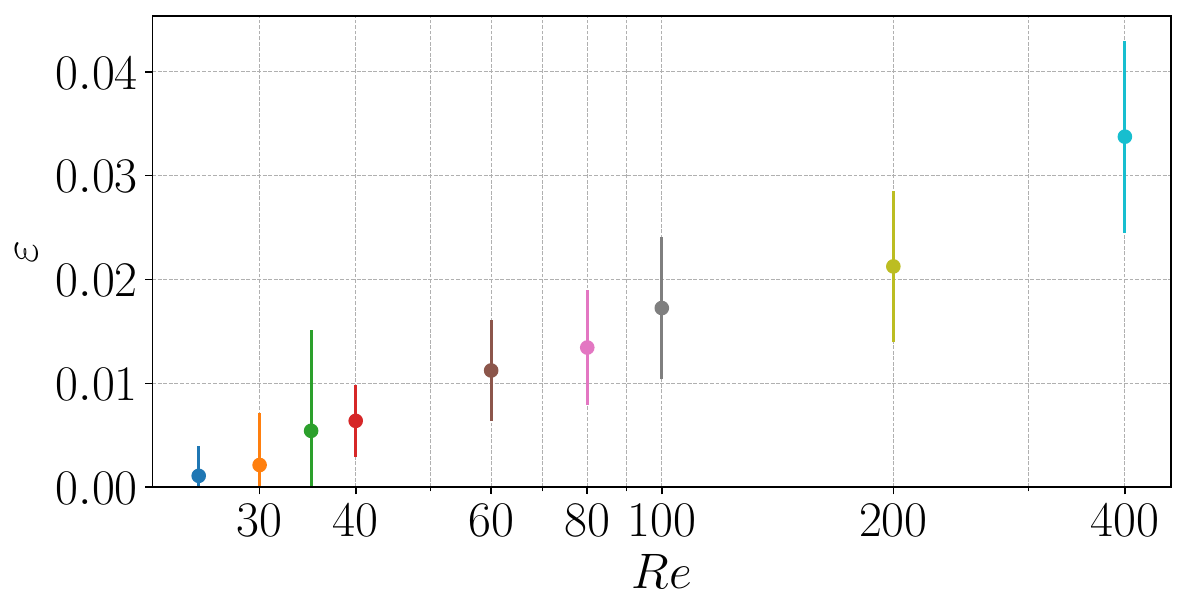}
    \caption{ The average reconstruction error $\varepsilon$ of the model against $Re$. \AC{A bottleneck dimension of $d_z = 1024$ is chosen for each $Re$.} The error bars at each $Re$ are defined by one standard deviation of the reconstruction error. \label{fig:error_all}}
\end{figure}

We report the relative error for each snapshot $\omega_j$ in the test datasets,
\begin{equation}
    \varepsilon_j \coloneq \frac{\| \omega_j - \mathscr{A}(\omega_j) \|}{\| \omega_j \|},
\end{equation}
where the norm is defined as $\| \omega_j \| \coloneq \sqrt{(1/4\pi^2) \iint \omega_j^2 d^2\mathbf{x}}$.
The average test error for all $Re$ examined is reported in figure \ref{fig:error_all}, with error bars defined by one standard deviation.
As expected, the error increases as $Re$ is increased, from $\varepsilon \sim 0.5 \%$ at $Re = 40$, up to $\varepsilon \sim 3.5 \%$ at $Re = 400$. 
These models are more than twice as accurate as those described previously in \cite{Page2024}.

\begin{figure}%[!ht]
    \includegraphics[width=\linewidth]{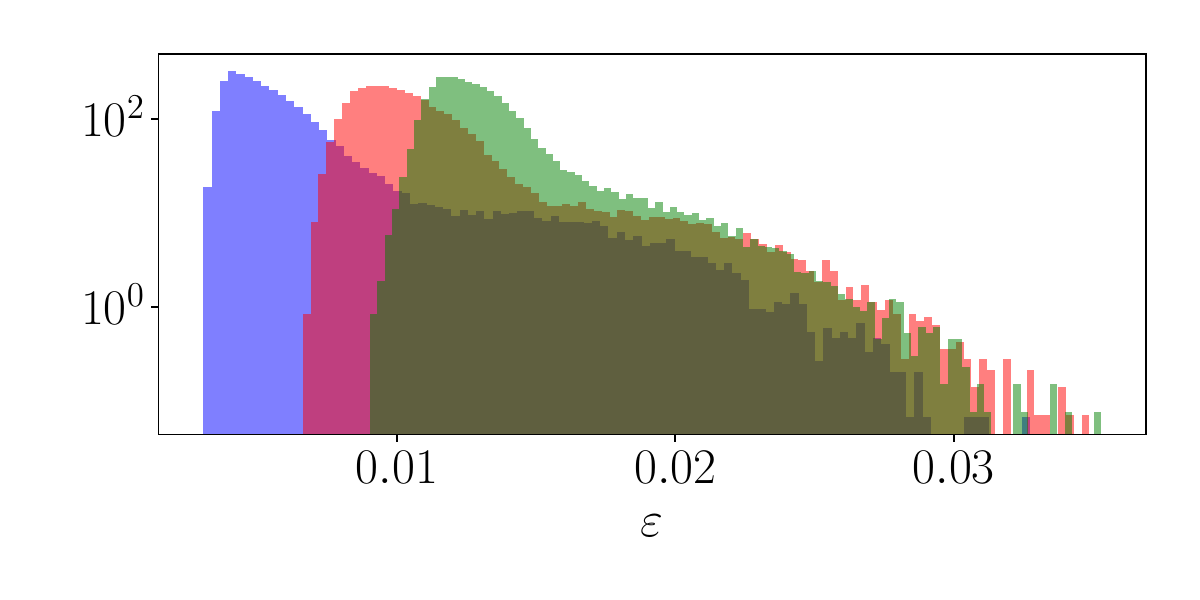}
    \caption{ The distribution of the reconstruction error $\varepsilon$ at $Re = 40$ is shown for the best model trained with data augmentation and no symmetry reduction (green), the best model from the first AdamW optimizer stage on symmetry reduced data (red), the best model from the second Adagrad optimizer stage on symmetry reduced data (blue). \label{fig:adagrad_symm}}
\end{figure}

The role of some of the new features introduced in the data pre-processing/network training approach is assessed in figure \ref{fig:adagrad_symm}.
% We report the improvement of the test error at $Re = 40$ due to the two-stage training process and the full symmetry reduction in figure \ref{fig:adagrad_symm}.
The distribution of errors from the best model from the first stage of training with the AdamW optimizer (red) is notably shifted compared with the best model from the second stage of training with the AdaGrad optimizer (blue).
Heuristically, this two-optimizer training process first explores the training loss landscape via the inherent momentum and bias correction of AdamW, before AdaGrad's feature-adapted learning rate allows it to settle more deeply into the best local minimum encountered in the first training stage. 
%The two-optimizer training was found to be most impactful at moderate $Re \lesssim 100$, with effectively negligible improvement for higher $Re$.
To quantify the impact of the full symmetry reduction preprocessing, we train on a dataset containing all symmetries in which data augmentation is applied at each training epoch by a random application of the discrete and continuous symmetry operations. 
Even with the data augmentation, the reconstruction error of this model (green) is notably larger than the model trained on fully reduced data, as shown in figure \ref{fig:adagrad_symm}. 
We shall also see in §\ref{sec:latent_structure} that explicitly factoring out the symmetries leads to a rich latent-space structure.

\begin{figure*}%[!ht]
    \centering
    \includegraphics[width=\linewidth]{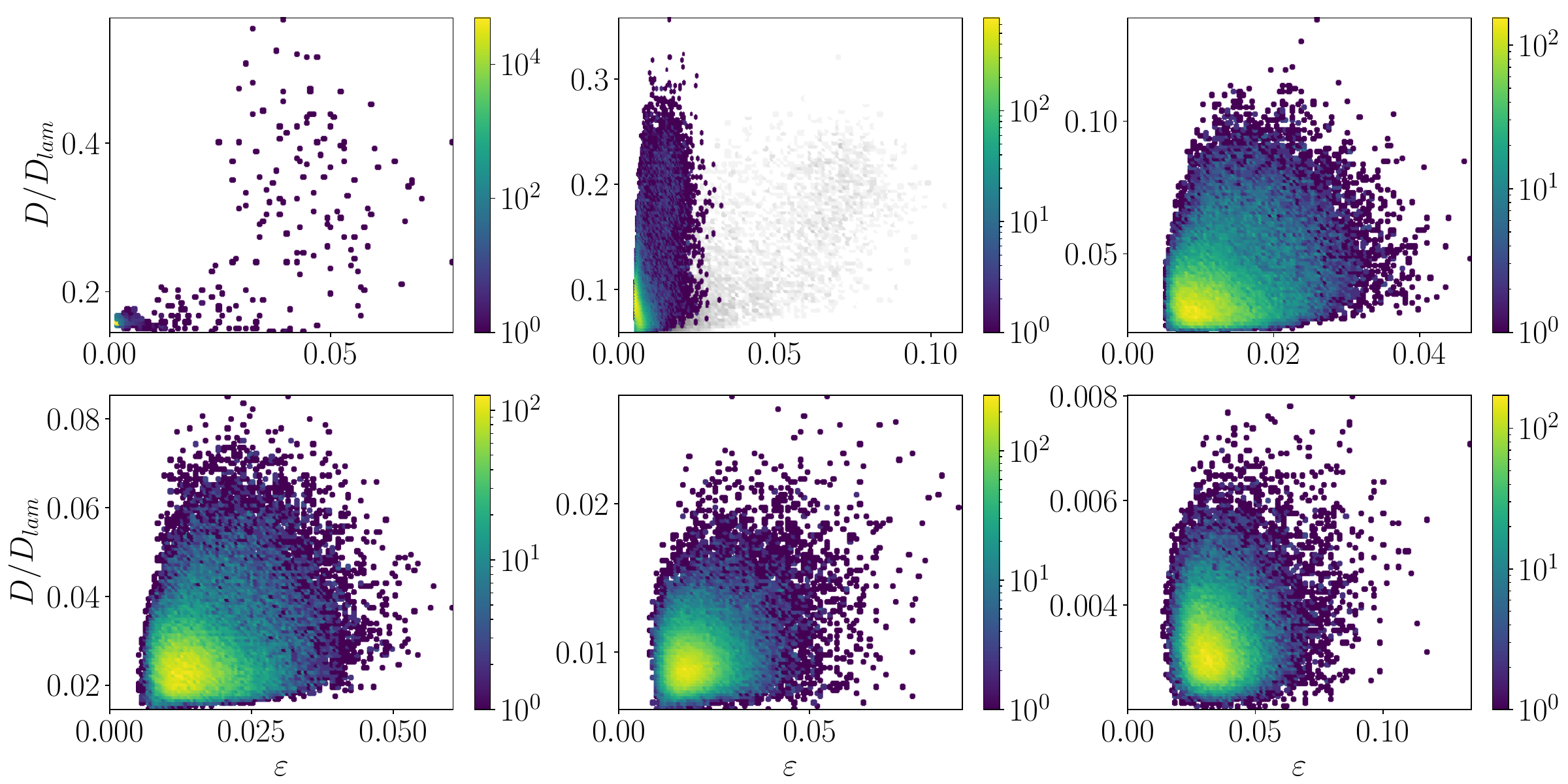}
    \caption{Histograms of the test datasets at (top) $Re = 25, 40, 80$ and (bottom) $Re = 100, 200, 400$ from left to right, comparing the snapshot dissipation rate $D$ normalized by the laminar dissipation rate $D_{lam} = Re/(2n^2)$ to the reconstruction error $\varepsilon$ of the snapshot. The colored histograms in each panel correspond to models trained with resampled training data, while the gray histogram at $Re=40$ (also on a logarithmic scale) corresponds to the model trained on the same number of non-resampled data. \label{fig:error_diss}}
\end{figure*}

The distribution of the error and correlation with dissipation rate is examined for a sample of the $Re$-values considered in figure \ref{fig:error_diss}.
For the $Re=40$ results, we also compare the performance of identical architectures trained on dissipation-resampled (color) or not resampled (gray) data at $Re = 40$. 
For a fair comparison, a similar number of training snapshots were used in both cases.
The resampling strategy considerably improves the error of the high dissipation events, and surprisingly also slightly improves the error of the low dissipation events. 
%Although the reconstruction error is still slightly correlated with dissipation rate, the resampling strategy and full symmetry reduction pre-processing of the data improves the reconstruction error compared to the models considered in PHBK.
The reconstruction error is still slightly correlated with dissipation rate, especially at higher $Re$.

\begin{figure}%[!ht]
    \includegraphics[width=\linewidth]{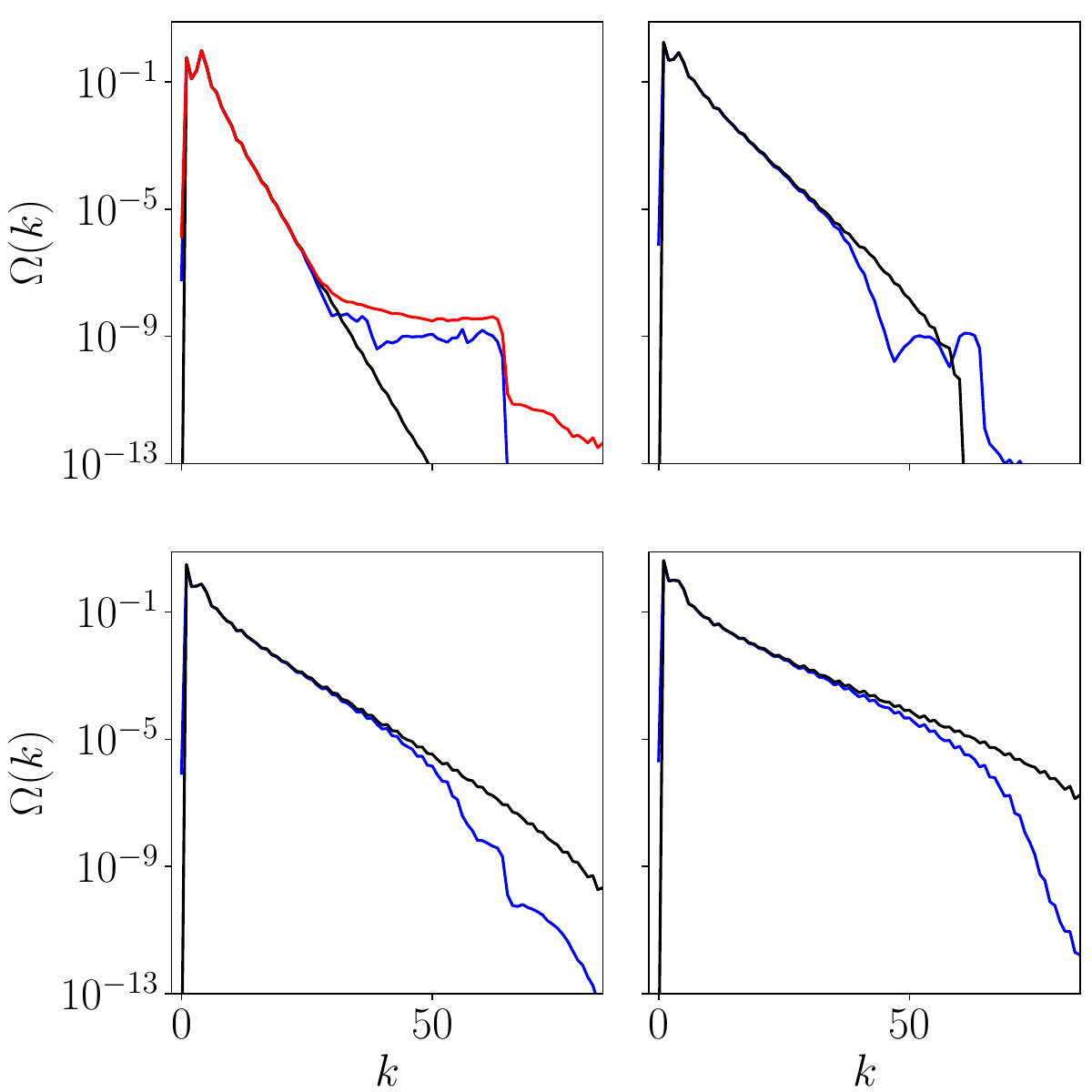}
    \caption{ The enstrophy spectrum $\Omega(k)$ of \AC{the (blue) reconstructed snapshots $\mathcal{A}(\omega)$ and (black) true snapshots $\omega$ at (top left) $Re = 40$, (top right) $Re = 100$, (bottom left) $Re = 200$ and (bottom right) $Re = 400$.}
    %The dotted blue lines indicate the spectrum of the intermediate layers within the output stream, becoming less opaque with each intermediate layer in the output stream. 
    The red line indicates the enstrophy spectrum at $Re = 40$ of a model trained with a single linear convolutional output layer of filter size $(8,8)$.\label{fig:spectrum}}
\end{figure}

We report the test-averaged \AC{enstrophy spectra $\Omega(k)$} generated by the model at $Re = 40, 100, 200$ and $400$ in figure \ref{fig:spectrum}.
%The ground truth from the training data is plotted in black, while the straight blue spectrum is the spectrum of the reconstructed data. 
We have replaced the traditional single output layer in our architecture by a sequence of linear convolutional layers of decreasing kernel sizes from $(16,16)$ to $(6,6)$ (full details in appendix \ref{app:arch_details}).
This `output stream' of layers was motivated by the empirical observation that the reconstructed \AC{enstrophy} spectrum with a single output convolutional layer plateaued incorrectly at high wavenumbers. 
This plateau is evident in \AC{the top left panel of} figure \ref{fig:spectrum}, where the red residual spectrum is generated by an identical fully-trained architecture with a single linear output layer of filter size $(8,8)$ instead of the output stream.
% The output stream was constructed such that the largest filters should model the large scale structures, which are progressively smoothed and refined as the filter sizes are reduced, allowing for a more accurate reconstruction across all spatial scales.
Empirically, the reduction in filter size through the stream tends to sequentially refine at finer spatial scales.
%The motivation for this output stream of layers is to allow the network to more accurately reconstruct the input across all spatial scales, such that the largest kernels should model the large scale structures, which become increasingly refined with the smaller kernels. 
%The dashed blue lines correspond to the output of the intermediate output layers within the output stream of linear convolutions of decreasing filter size. 
%The dashed lines get progressively darker from the first output layer of filter size $(16,16)$, through to the second last output layer of filter size $(6,6)$ (full details in appendix \ref{app:arch_details}).
%We have normalized the spectra such that $E(k = n) = 1$ to enable a comparison of spatial scales across layers.
%As expected, the small spatial scales are increasingly refined throughout this output stream.
Comparing the \AC{enstrophy} spectra of the output with and without the output stream at $Re = 40$, we see a notable improvement at the largest spatial scale, and a slight improvement at the very small spatial scales.
We found that using any smaller filter sizes resulted in an over-smoothing of the reconstructed vorticity fields.
\AC{This plateau at the high wavenumbers is increasingly absent at higher $Re$, and virtually non-existent at $Re = 400$. 
Instead, the reconstructed fields are overly smoothed at the higher $Re$ values of $200$ and $400$.}
%As mentioned previously, the spurious energy plateau in the high spatial scales was common to all models trained, and could not be fully overcome using the proposed output stream architecture. 

\subsection{Attractor dimension}

Here we present the main result of this study -- the scaling with $Re$ of the estimated dimension of the chaotic attractor.
Implicit rank minimization drives the rank of the embedding space to an approximately minimal rank solution, which we use as (an upper bound for) our numerically estimated dimension.
To estimate this minimal rank, we perform a singular value decomposition (SVD) on the \AC{latent data matrix $\mathbf Z \mathbf{Z}^T$ \citep{Zeng_2024}, where $\mathbf{Z} = [\dots z_i \dots ]$ and $z_i = \mathscr W (\mathscr E(\omega_i))$ for all $\omega_i$ in the test dataset.}
The SVD of $\mathbf Z \mathbf Z^T = \mathbf U \boldsymbol\Sigma \mathbf U^T$ 
% [Should there be a square here? The SVD should be of Z itself, correct?]} 
computes the singular vectors $\mathbf u_j \in \mathbb{R}^{d_z}$ where $\mathbf U = [\mathbf u_0 \dots \mathbf u_{d_{z-1}}]$ (the left singular vectors of $\mathbf Z$) and singular values $\sigma_j \in \mathbb{R}$ where $\boldsymbol \Sigma = \text{diag}(\sigma_0, \dots, \sigma_{d_{z-1}})$ (the square of the singular values of $\mathbf Z$).
The number of non-negligible singular values in this spectrum is the estimate for $d_{\mathcal{A}}$ in the fully symmetry reduced subspace.
% Each singular vector corresponding to a non-negligible singular value is a necessary direction, or degree of freedom, in representing the original snapshot in the latent space.
Each of these singular vectors represents some important structure of the original data in the latent space, which will be investigated in more detail in §\ref{sec:latent_structure}.

\begin{figure}%[!ht]
    \includegraphics[width=\linewidth]{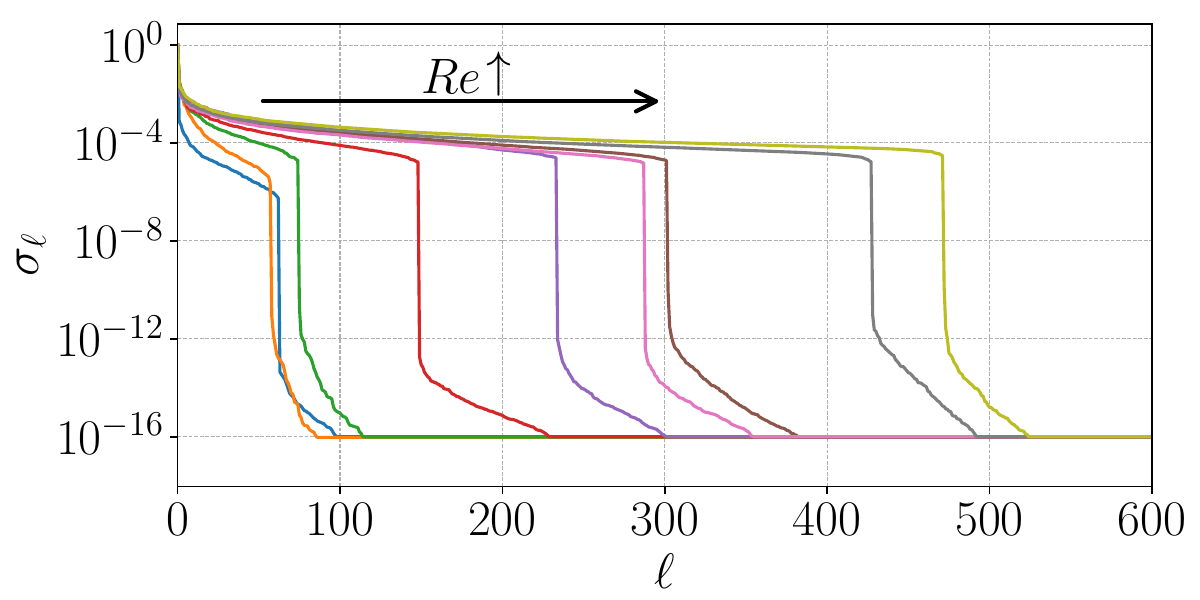}
    \includegraphics[width=\linewidth]{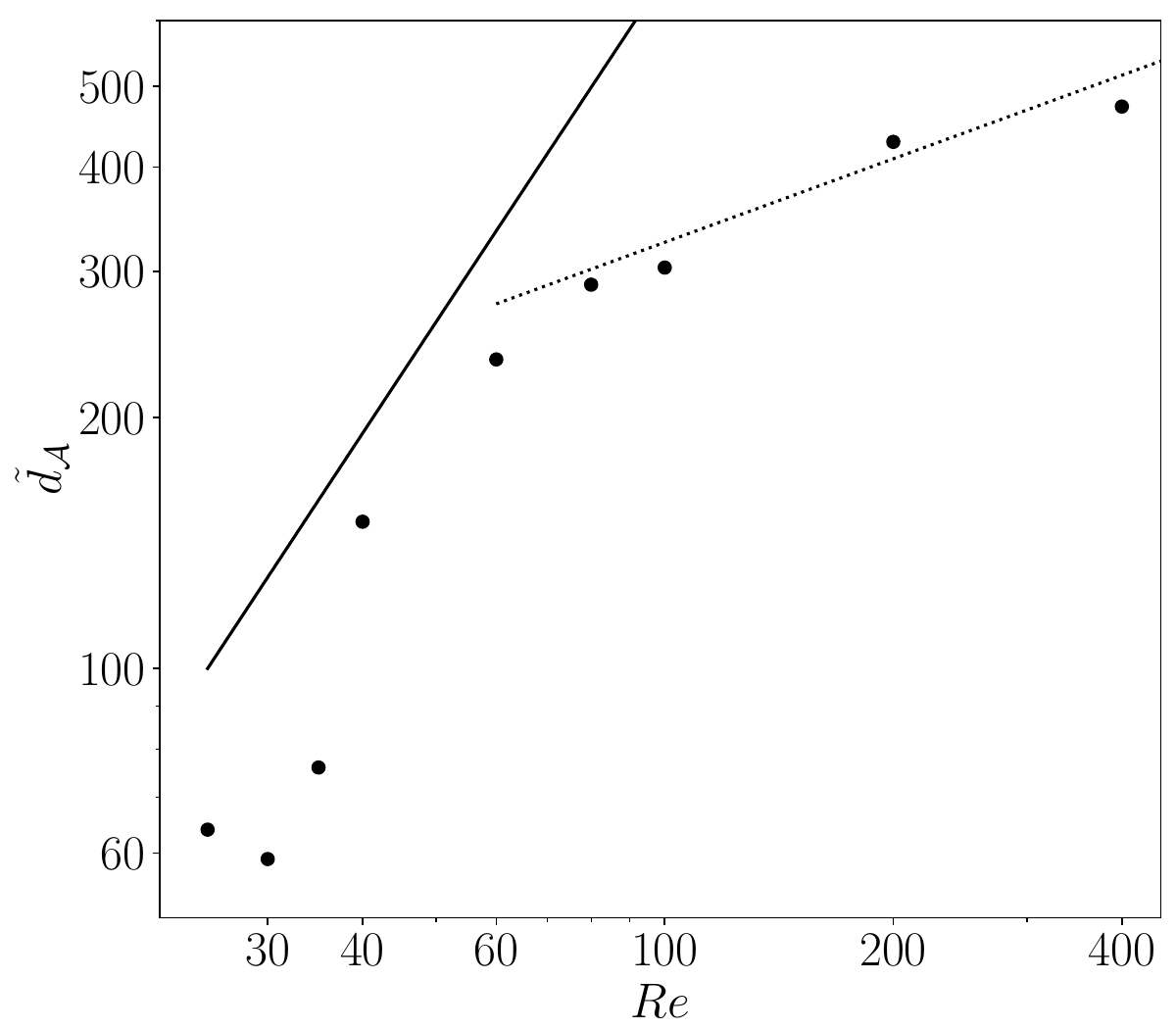}
    \caption{(Top) Singular values $\sigma_i$ in the singular value decompositions of the covariance of the latent data matrix at $Re =$ 25 ({\color{rea} $\lline$}), 30 ({\color{reb}$\lline$}), 35 ({\color{rec}$\lline$}), 40 ({\color{red}$\lline$}), 60 ({\color{ree}$\lline$}), 80 ({\color{ref}$\lline$}), 100 ({\color{reg}$\lline$}), 200 ({\color{reh}$\lline$}) and 400 ({\color{rei}$\lline$}). (Bottom) The estimated upper bound for the dimension of the turbulent attractor $\Tilde{d}_{\mathcal{A}}$ against $Re$. The bold line shows the rigorous upper bound scaling $\sim Re^{4/3}(1 + \log{(2^{5/2}\pi^3 Re^2)})^{1/3}$ and the dotted line shows the scaling $\sim Re^{1/3}$.  \label{fig:singvals_dim}}
\end{figure} 

The singular values for each $Re$ considered are reported in \AC{the top panel of figure \ref{fig:singvals_dim}}.
In all cases there is a sharp drop before $\sigma_j$ becomes negligibly small at a critical index $j = c$ \AC{(defined as the first index where $\sigma_{c} < 10^{-8}$)}, corresponding to the rank of the learned, fully symmetry-reduced, latent space. 
\AC{To compute the estimate for $d_{\mathcal{A}}$ in full state space, which we denote by $\Tilde{d}_{\mathcal{A}}$, we add the continuous symmetry back into the estimate, such that $\Tilde{d}_{\mathcal{A}} = c + 1$ (reducing the discrete symmetries reduces the size of the state space, but not the dimension).} 
%We trained two models at each $Re$, resulting in two independent estimates for $\Tilde{d}_{\mathcal{M}}$ which agreed very closely
We computed two estimates for $\Tilde{d}_{\mathcal{A}}$ at each $Re$ by training two randomly initialized autoencoders from scratch.
These estimates were found to be robust to the training -- they differed only slightly between models at all $Re$. %, suggesting a robustness to these estimates.
The smallest estimate at each $Re$ is reported in \AC{the bottom panel of figure \ref{fig:singvals_dim}}.

% discussion of the analytic scaling and relation of G to Re. 
The non-dimensional Grashof number $G$ in the rigorous upper bound \citep{CONSTANTIN1988, DOERING1991} of the \emph{global} attractor dimension is related to the present definition of $Re$, with monochromatic $n=4$ Kolmogorov forcing, by $G = 2^{5/2} \pi^3 Re^2$. 
This upper bound then scales with $Re$ like $b Re^{4/3}(1 + \log{(2^{5/2}\pi^3Re^2)})^{1/3}$, and is indicated by the bold line in \AC{the bottom panel of figure \ref{fig:singvals_dim}} ($b$ is an unspecified constant).
% A heuristic estimate for the dimension provided in} \citep{ohkitani1989} scales as $\sim Re^{4/3}(\log (2^{5/2}\pi^3Re^2))^{1/3}$, similarly to the rigorous bound to leading order}. % don't think we need this now given our discussion of global/turbulent attractor
This bound and the lower bound in \citep{Liu1993} \AC{for the global attractor} both scale with the number of degrees of freedom in a two-dimensional domain ($Re^{4/3}$ is proportional to the square of the domain size $L$ over the Kraichnan scale) with a logarithmic correction.

In contrast, our data-driven estimate for the dimensionality of the \AC{\emph{chaotic}} attractor explored by long turbulent trajectories exhibits a much weaker scaling at high $Re$.
While we observe some non-monotonic behavior in \AC{the bottom panel of figure \ref{fig:singvals_dim}} at lower $Re$, there is a sharp transition beyond $Re=40$ beyond which an approximate scaling $d_{\mathcal A} \sim Re^{1/3}$ is observed \AC{in the range $Re \in [60, 400]$} -- identified with the dotted line in the figure.
\JP{Although this scaling was estimated over this limited range of $Re$ values, earlier work \cite{Chandler2013,cleary2025} indicates that an `asymptotic' regime (a clear power law scaling of dissipation with $Re$) is established for $Re\gtrsim 60$ and extends to at least $Re=1000$, which would suggest that our dimension scaling may well also continue to be robust beyond $Re=400$.}
% \AC{Although this scaling was only computed over this range of $Re$, \citep{cleary2025} and \citep{Chandler2013} present evidence of an `asymptotic regime' of two-dimensional Kolmogorov flow, which also begins at $Re \sim 60$ and extends at least to $Re \sim 1000$.}
\AC{The estimate $d_{\mathcal A} \sim Re^{1/3}$ is a weaker scaling than a previous result of $\sim Re^{0.78}$, which was computed via the Kaplan-Yorke conjecture \citep{clark2020}.
However, this flow configuration was forced by a different profile and was damped at large scales, making a direct comparison challenging.  
}

\begin{figure}%[!ht]
    \centering
    \includegraphics[width=\linewidth]{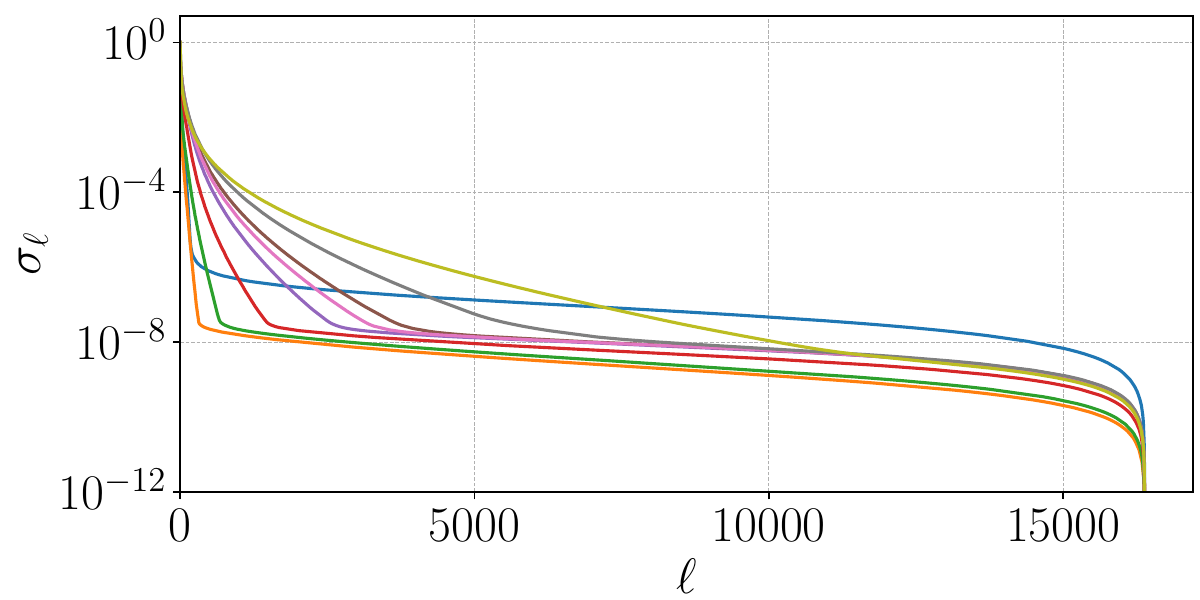}
    \caption{\AC{Principal component analysis values $\sigma_{\ell}$ for the symmetry-reduced test datasets at $Re =$ 25 ({\color{rea} $\lline$}), 30 ({\color{reb}$\lline$}), 35 ({\color{rec}$\lline$}), 40 ({\color{red}$\lline$}), 60 ({\color{ree}$\lline$}), 80 ({\color{ref}$\lline$}), 100 ({\color{reg}$\lline$}), 200 ({\color{reh}$\lline$}) and 400 ({\color{rei}$\lline$}).}  \label{fig:pca_vals}}
\end{figure}

\AC{\JP{As a point of comparison for} the non-linear dimension estimation by the IRMAE networks, we \JP{also performed} a linear dimensionality reduction approach on the symmetry-reduced test datasets using principal component analysis (PCA) \citep{Pearson01111901}.}
\AC{The \JP{singular values from a PCA performed} at each $Re$ are reported in figure \ref{fig:pca_vals}.}
\AC{There is no sharp cut-off in the singular values, especially at the higher $Re$, so it is not possible to meaningfully extract a dimension at each $Re$ using this approach.}
% We find that the dimension estimate increases sharply beyond $Re = 40$, coinciding with the transition to the fully turbulent regime. 
 % similar sharp transition point was reported in Rayleigh-Bénard flow, at a Rayleigh number of $Ra \sim 10^7$, which also coincides with the transition to turbulence in this flow. 
% Beyond this transition point, we report an approximate scaling of the dimension as $Re^{1/3}$, indicated by the dotted line in figure \ref{fig:singvals_dim}.
% below is a flow with large-scale damping and so is not relevant
% This estimate also} has a weaker scaling than the previous numerical estimates of $\sim Re^{0.78}$ computed via the Kaplan-Yorke conjecture \citep{clark2020}. %, and is consistent with the analytic upper bound.
% Interestingly, the heuristic scaling provides a reasonable fit to our estimates over the moderate $Re \lesssim 100$ considered.  
% The estimates at $Re < 40$ are likely not monotonic with $Re$ as the data in this meta-stable regime is dominated by some attracting limit cycle (see the distribution of dissipation rate in top left panel of figure \ref{fig:error_diss}), affecting the estimated rank of the learned embedding.  

One caveat with our estimates is that our models do not perfectly reconstruct their inputs, and this error does increase with $Re$. 
However, the markedly weaker scaling with $Re$ is consistent with the formation of coherent structures which are associated with a subset of wavenumber pathways in a Fourier representation.
\JP{We also examined the relative error with respect to a reference \emph{trajectory} as the model predictions are unrolled in time and found that the errors remain roughly constant for longer than a Lyapunov time (we compared at $Re=100$ where this time has been estimated at $T_{\lambda} = 3.6$ by \cite{velamartin2024}).}
\AC{Another caveat is we also assume that we have sufficiently sampled the chaotic attractor with our test datasets of 50,000 snapshots at each $Re$.}
We now explore the representation of the turbulence learned in these models by decoding individual components from the SVD.

\section{Latent space structure}
\label{sec:latent_structure}

In this section, we investigate the structure of the latent space learned by the model at various values of $Re$. 
By explicitly factoring out the symmetries of Kolmogorov flow, the model does not have to learn implicit representations of these symmetry groups. 
This results in a much richer representation of the dynamics in the fundamental symmetry chart, which can be explicitly related to all symmetric copies.

\subsection{Modal structures}

We first examine the contribution to the vorticity field reconstruction from each of the singular vectors (modes) \AC{$\mathbf u_{\ell}$} in the SVD of $\mathbf Z \mathbf Z^T$.
% [Again here -- I think you mean the eigendecomposition of $\mathbf Z \mathbf Z^T$, which is related to the SVD of $\mathbf Z$, correct? Please update accordingly, and also note that if you are finding eigenvalues then these are the square of the singular values...]}.
To do this, we project low and high dissipation snapshots onto the $\ell = 0$ mode and combinations of this mode with higher modes.
The projection operator onto mode $\ell$ in the embedding space for some vorticity field $\omega$ is defined by 
\begin{equation}
    \mathcal{P}_{\ell}(\omega) = \langle \mathbf u_{\ell}, \mathscr{W}(\mathscr{E}(\omega)) \rangle,
\end{equation}
where $\langle \cdot , \cdot \rangle : \mathbb{R}^{d_z} \times \mathbb{R}^{d_z} \to \mathbb{R}$ denotes the inner product in the embedding space.
The projection onto the non-negligible modes $\ell < c$ defines the co-ordinates of $\omega$ in minimal-rank co-ordinates
\begin{equation}
    \phi(\omega) \coloneq (\mathcal{P}_{0}(\omega), \dots, \mathcal{P}_{c-1}(\omega) ) \in \mathbb{R}^{c}.
    \label{eq:tsne_in}
\end{equation}

\begin{figure*}%[!ht]
    \includegraphics[width=0.5\linewidth]{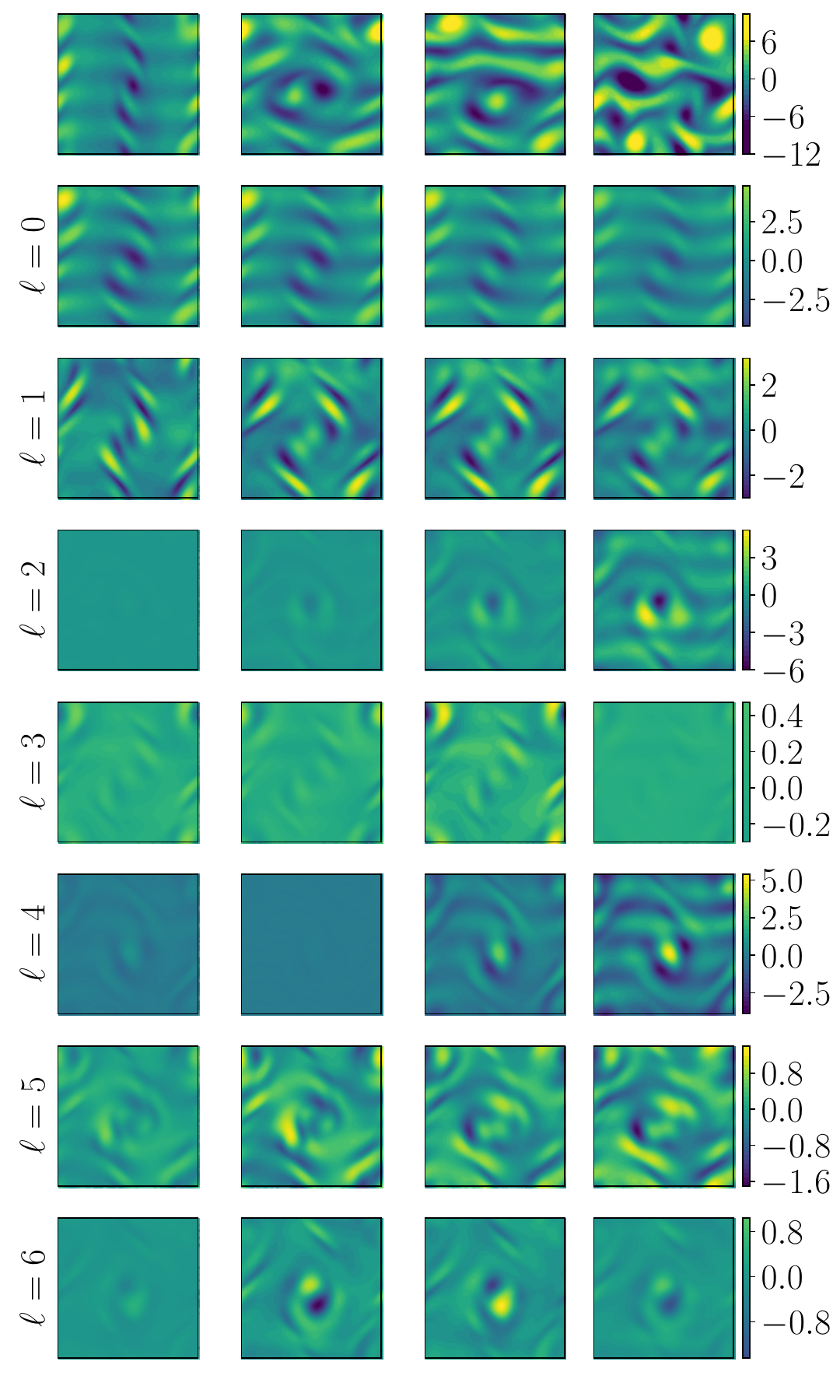}%
    \includegraphics[width=0.5\linewidth]{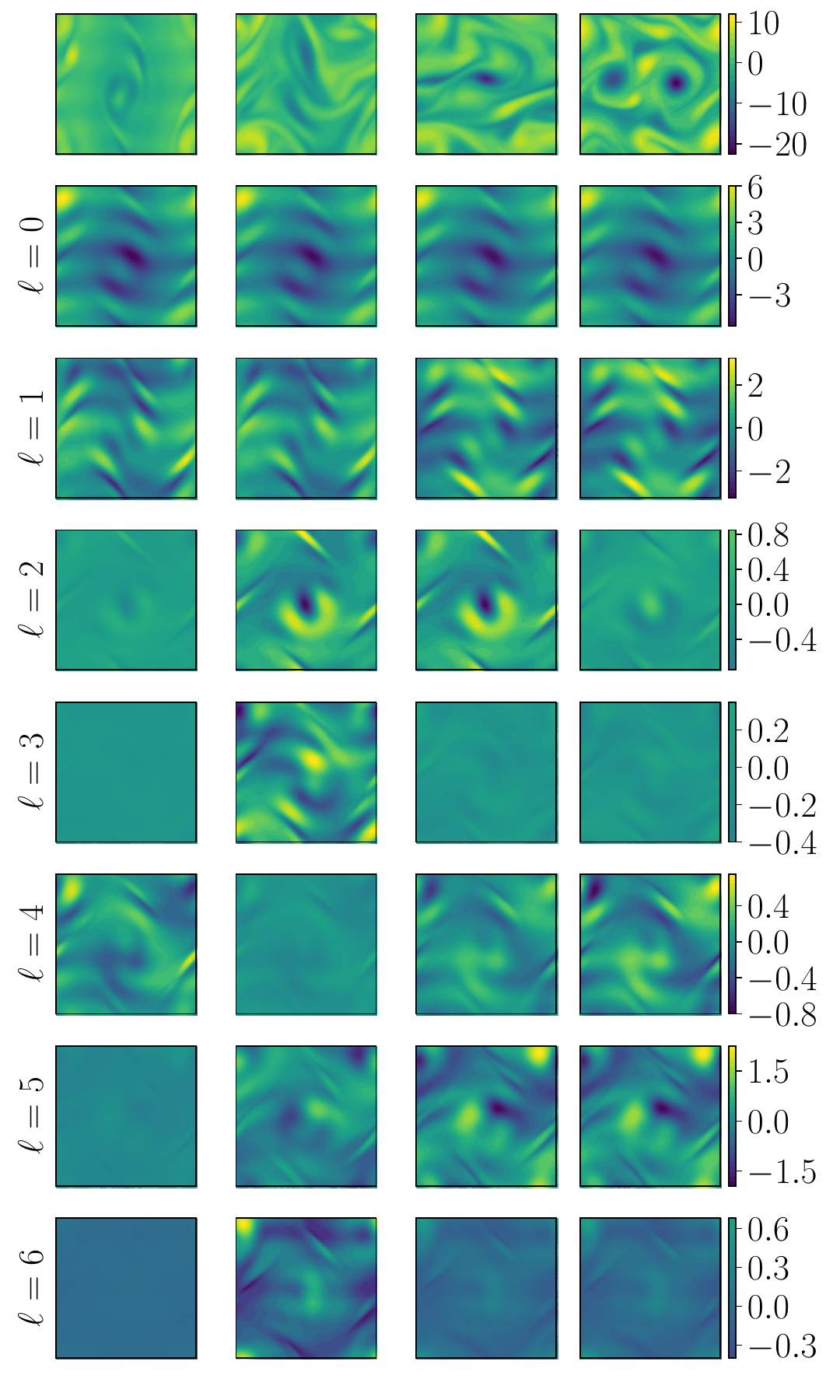}
    \includegraphics[width=0.5\linewidth]{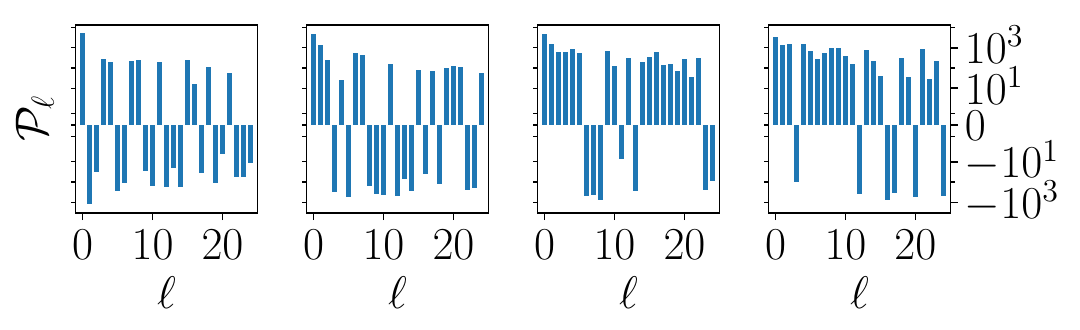}%
    \includegraphics[width=0.5\linewidth]{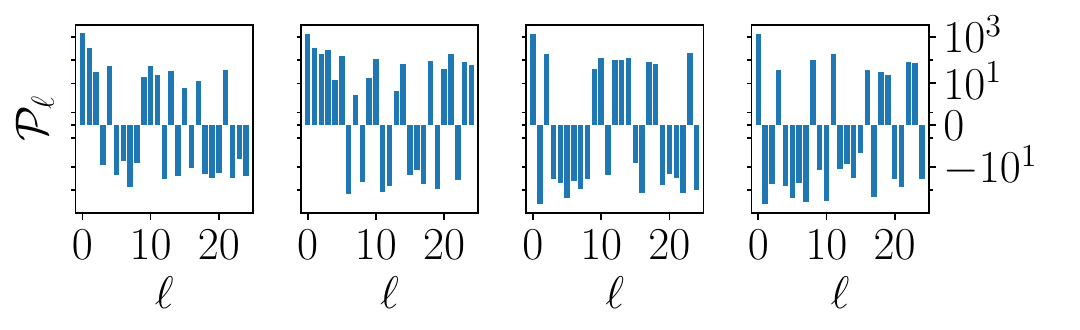}
    \caption{The contributions $\mathcal{D}(\mathcal{P}_0(\omega)\mathbf u_{0})$ from mode 0 and $\mathcal{D}(\mathcal{P}_{\ell}(\omega)\mathbf u_{\ell} + \mathcal{P}_0(\omega)\mathbf u_{0}) - \mathcal{D}(\mathcal{P}_0(\omega)\mathbf u_0)$ from modes $\ell = 1, \dots, 6$. The top row shows the input vorticity snapshots at (left panels) $Re = 40$ with $D / D_{lam} = 0.06, 0.12, 0.18, 0.26$ from left to right, (right panels) $Re = 100$ with $D / D_{lam} = 0.016, 0.032, 0.047, 0.064$ from left to right. Each column shows the contribution from mode $\ell$ to the reconstruction of the input vorticity snapshot. Each snapshot in a row shares the same colorbar to highlight the activations of different modes across dissipation rates. \AC{The bottom row shows the projections $\mathcal{P}_{\ell}$ (on a logarithmic scale) onto the first 25 latent modes for the snapshot in each column.} \label{fig:modes}}
\end{figure*}
\begin{figure*}%[!ht]
    \includegraphics[width=0.5\linewidth]{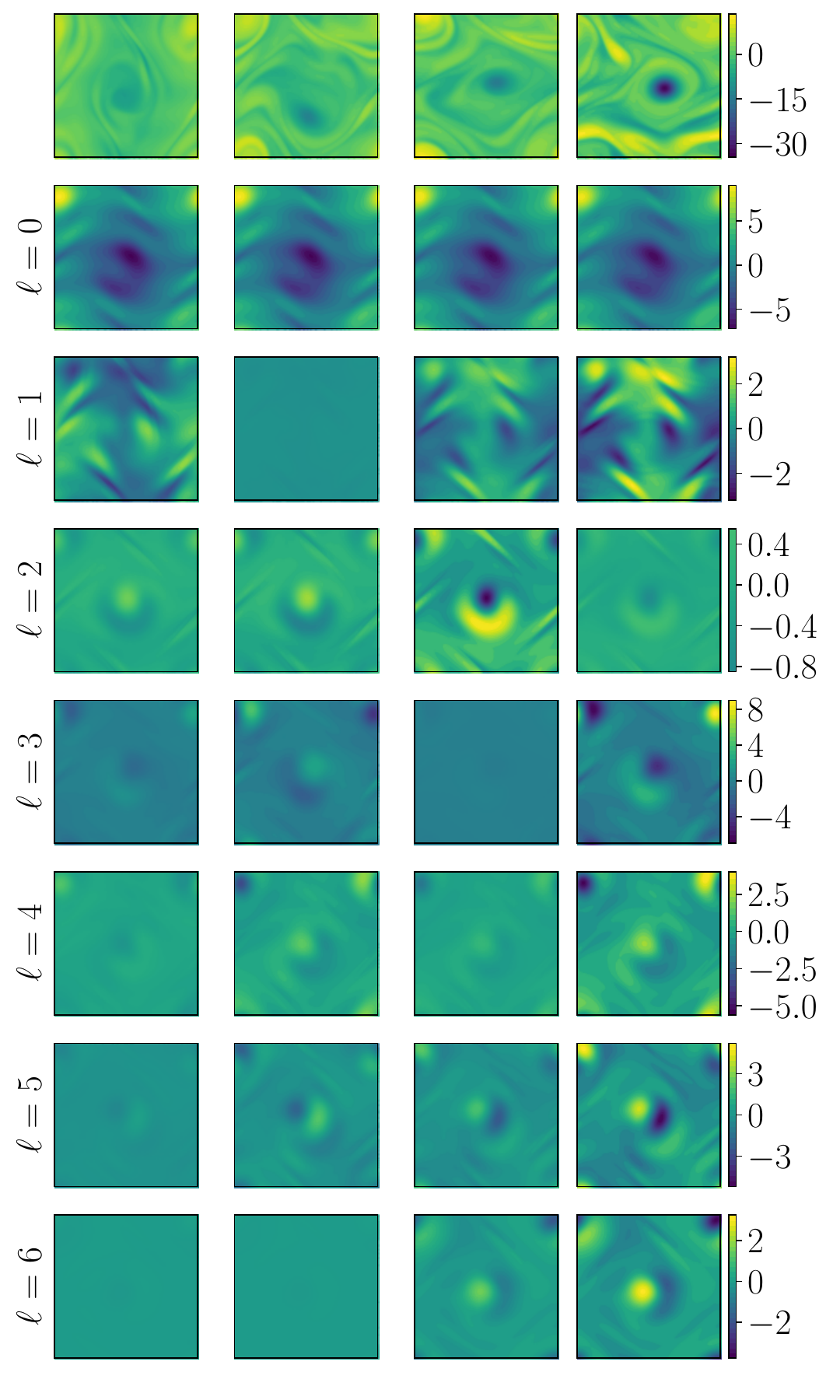}%
    \includegraphics[width=0.5\linewidth]{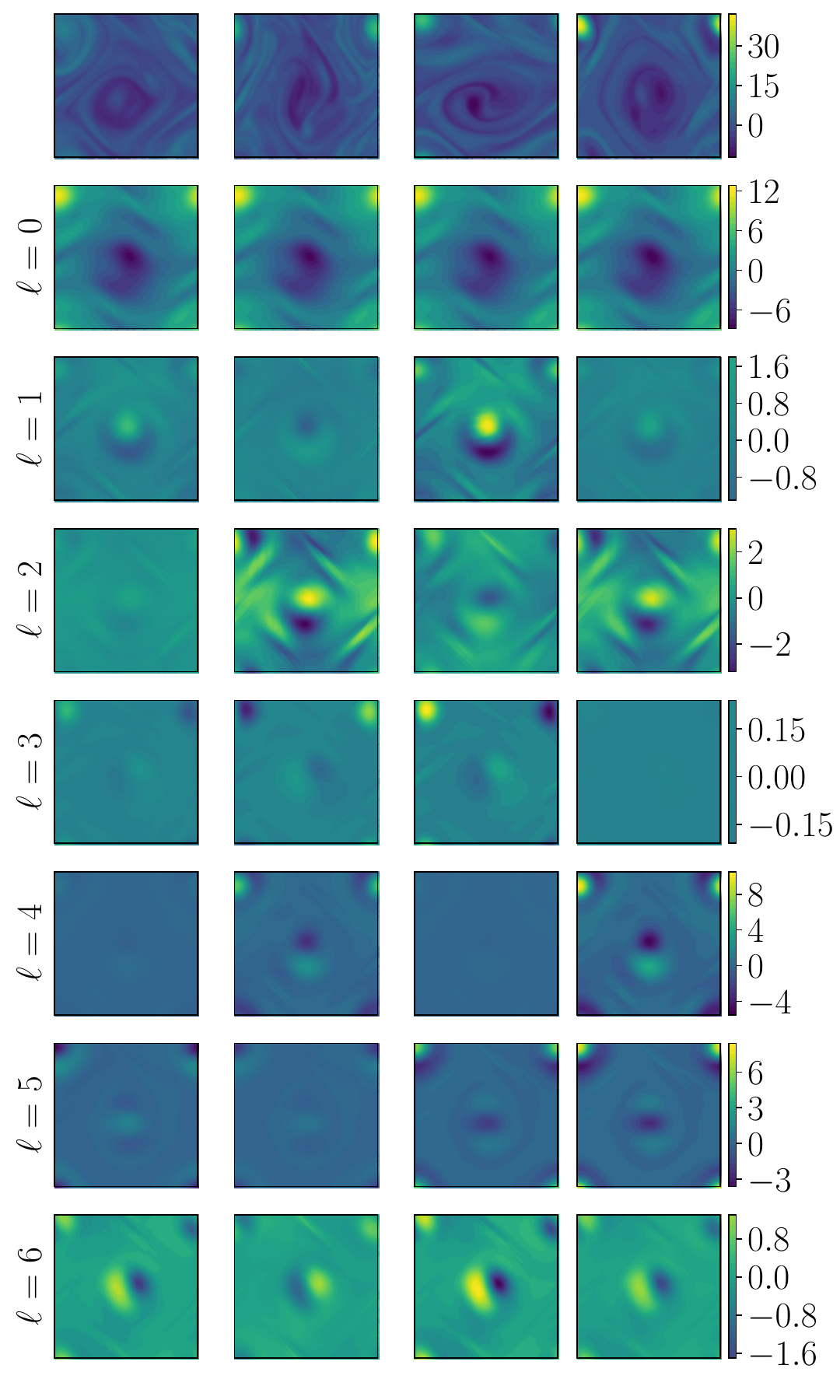}
    \includegraphics[width=0.5\linewidth]{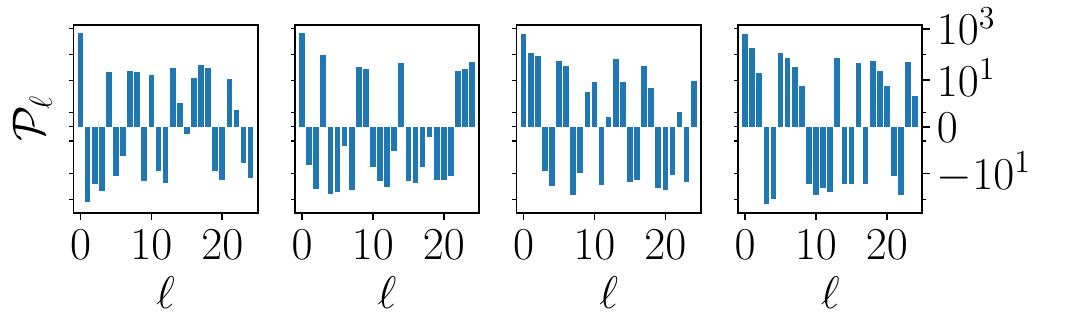}%
    \includegraphics[width=0.5\linewidth]{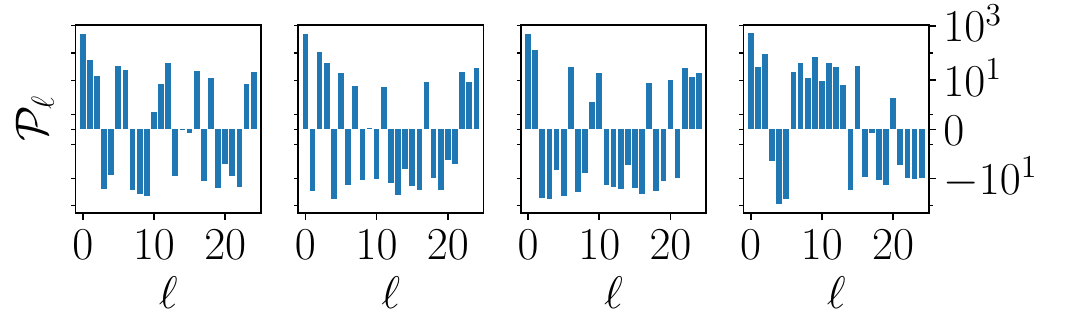}
    \caption{The contributions $\mathcal{D}(\mathcal{P}_0(\omega)\mathbf u_{0})$ from mode 0 and $\mathcal{D}(\mathcal{P}_{\ell}(\omega)\mathbf u_{\ell} + \mathcal{P}_0(\omega)\mathbf u_{0}) - \mathcal{D}(\mathcal{P}_0(\omega)\mathbf u_0)$ from modes $\ell = 1, \dots, 6$. The top row shows the input vorticity snapshots at (left panels) $Re = 200$ with $D / D_{lam} = 0.0064, 0.0088, 0.0102, 0.0233$ from left to right, (right panels) $Re = 400$ with $D / D_{lam} = 0.0022, 0.003, 0.0035, 0.0054$ from left to right. Each column shows the contribution from mode $\ell$ to the reconstruction of the input vorticity snapshot. Each snapshot in a row shares the same colorbar to highlight the activations of different modes across dissipation rates. \AC{The bottom row shows the projections $\mathcal{P}_{\ell}$ (on a logarithmic scale) onto the first 25 latent modes for the snapshot in each column.}\label{fig:modes_highRe}}
\end{figure*}

To visualize the physical structures spanned by each mode, we decode the projections onto each mode back to physical space. 
We add the projection onto the $\ell = 0$ mode when projecting onto $\ell > 0$ modes \AC{i.e.\ $\mathcal{D}(\mathcal{P}_{\ell}(\omega)\mathbf u_{\ell} + \mathcal{P}_0(\omega)\mathbf u_{0})$} as otherwise the output $\mathcal{D}(\mathcal{P}_{\ell}(\omega)\mathbf u_{\ell})$ is not physically realistic \citep[for a similar discussion, see][]{Page2021, Page2024}.
Then, we subtract $\mathcal{D}(\mathcal{P}_0(\omega)\mathbf u_{0})$ from the decoded field, in an effort to isolate and interpret the contribution from the $\ell > 0$ mode in physical space.

The contributions $\mathcal{D}(\mathcal{P}_0(\omega)\mathbf u_{0})$ from mode 0 and $\mathcal{D}(\mathcal{P}_{\ell}(\omega)\mathbf u_{\ell} + \mathcal{P}_0(\omega)\mathbf u_{0}) - \mathcal{D}(\mathcal{P}_0(\omega)\mathbf u_{0})$ from modes $\ell = 1, \dots, 6$ are shown in figure \ref{fig:modes} for a selection of snapshots spanning the full range of dissipation rates at $Re = 40$ and $Re = 100$. 
The same modes for $Re = 200$ and $Re = 400$ are reported in figure \ref{fig:modes_highRe}.
At all $Re$ considered, the $\ell = 0$ mode bears some resemblance to the time-averaged vorticity state in the symmetry-reduced subspace across the full range of dissipation rates.
This structure is also visually similar to many low-dissipation \emph{exact} solutions in this flow (traveling waves and relative periodic orbits, see \cite{Chandler2013,lucas2015,Page2021} -- some examples at $Re=40$ are also included in appendix \AC{\ref{app:unstable}}).
There is a weaker projection onto this state for the high dissipation snapshots at $Re = 40$, whereas the projection at the higher $Re \in \{ 100, 200, 400\}$ is approximately equal across all snapshots presented.
%all the way through to a state closely resembling the equilibrium which bifurcates off the laminar state by a continuous-symmetry-breaking at low $Re \approx 10$ \citep[][]{Chandler2013, Lucas_Kerswell_2014}.
%It is interesting that, in the latent space, there exists a linear relationship between this mean flow state and this non-trivial equilibrium -- the chaotic dynamics eventually emerge via further bifurcations from this equilibrium.
Similar to previous studies \citep{Page2021, Page2024}, the $\ell = 0$ mode is central to the latent representation -- the singular value $\sigma_0$ is two orders of magnitude larger than $\sigma_1$ at $Re = 40$. 
The $\ell = 1$ mode also captures similar structures at both $Re= 40$ and $Re = 100$, and is dominated by paired diagonal elongated vortex cores of opposite circulation. 
This mode is visually similar to the most unstable eigenfunction of many low-dissipation exact solutions which resemble the mean flow state (see figure \ref{fig:TW_stab} in appendix \AC{\ref{app:unstable}}).
% AC: \ell = 1 mode does seem to be rotationally symmetric, I'm not sure about \ell = 0?
%Both $\ell = 0$ and $\ell =1$ modes appear to be rotationally symmetric.

At the higher $Re = 200$ and $Re = 400$, the representation changes somewhat -- there is less evidence of the Kolmogorov forcing profile in the decodes of the dominant SVD vectors.
Instead, many of the modes capture local features in the vorticity, including dipole/tripolar structures in different orientations (in particular note modes beyond $\ell \geq 3$.
%this structure is increasingly captured by the $\ell = 2$ mode, perhaps indicating that the structure is closely related to the Kolmogorov forcing profile, which becomes less dynamically important as $Re$ is increased.
%The mode $\ell = 1$ is quite similarly activated across all dissipation rates at $Re = 40$, likely closely related to the Kolmogorov forcing profile. 
%At $Re = 100$, the $\ell = 0$ mode instead resembles the mean flow state of the symmetry reduced subspace for all the snapshots sampled uniformly across the range of dissipation rates, while the $\ell = 1$ mode is instead more strongly activated for the high dissipation snapshots.

One striking aspect of the decodes at all $Re$ is the sparsity in the representation -- some modes are not active for the input snapshots.
\AC{For example, consider modes $\ell = 1$ and $\ell = 2$ in the left column of figure \ref{fig:modes}. 
The projections onto these modes $\mathcal{P}_{\ell}$ is plotted on a logarithmic scale in the bottom row of figure \ref{fig:modes}, demonstrating that $\mathcal{P}_2$ is two orders of magnitude smaller than $\mathcal{P}_1$.}
% At all $Re$ considered, there is a stark activation of specific modes for the high dissipation snapshots, in particular for $\ell = 2,4,5$ at $Re = 40$, resulting in a sparse representation of each snapshot in the basis of modes. 
This is suggestive of a strong separation of modes for different dynamical processes, but is also likely due in part to the discrete symmetry reduction.
The mapping to a fundamental chart is based on the first Fourier mode in the vertical. 
This can mean that snapshots on a trajectory can suddenly go through a `jump' to another chart which suddenly shifts the local vortical features when symmetry reduction is performed (the first Fourier mode may not be apparent by eye, particularly in high dissipation snapshots), and the autoencoder therefore must learn to represent small scale vortices in different areas of the symmetry-reduced snapshots. This is perhaps apparent in the $\ell = 3$ mode at $Re=400$ (figure \ref{fig:modes_highRe}) with a dipole in the upper corners of the decode, though in general the features across the modes reported in figures \ref{fig:modes} and \ref{fig:modes_highRe} are all physically distinct.
In contrast to similar attempts to visualize physical representations of individual latent directions (e.g.\ in \cite{Page2024}) the decodes are all smooth and physically interpretable, even at high $Re$.

\subsection{Low-dimensional visualizations}
\label{sec:low-d-v}

\begin{figure*}
    \centering
    \begin{minipage}{0.5\linewidth}
        \centering
        \includegraphics[width=\linewidth]{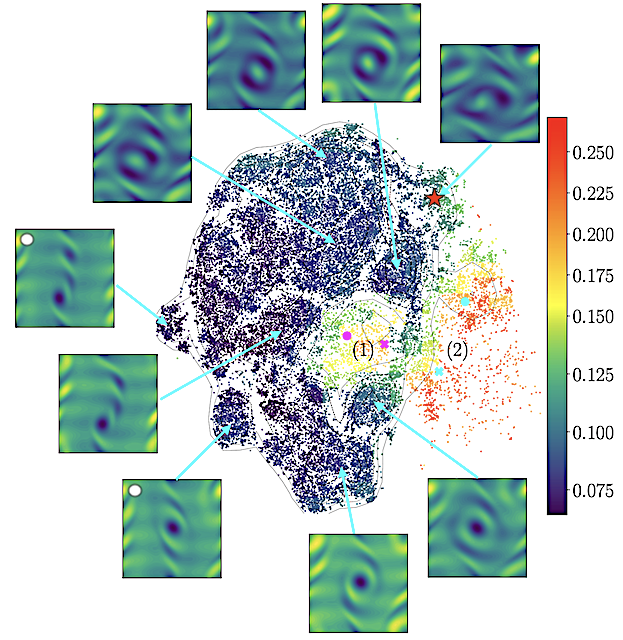}
    \end{minipage}%
    \vrule
    \begin{minipage}{0.5\linewidth}
        \centering
        \includegraphics[width=\linewidth]{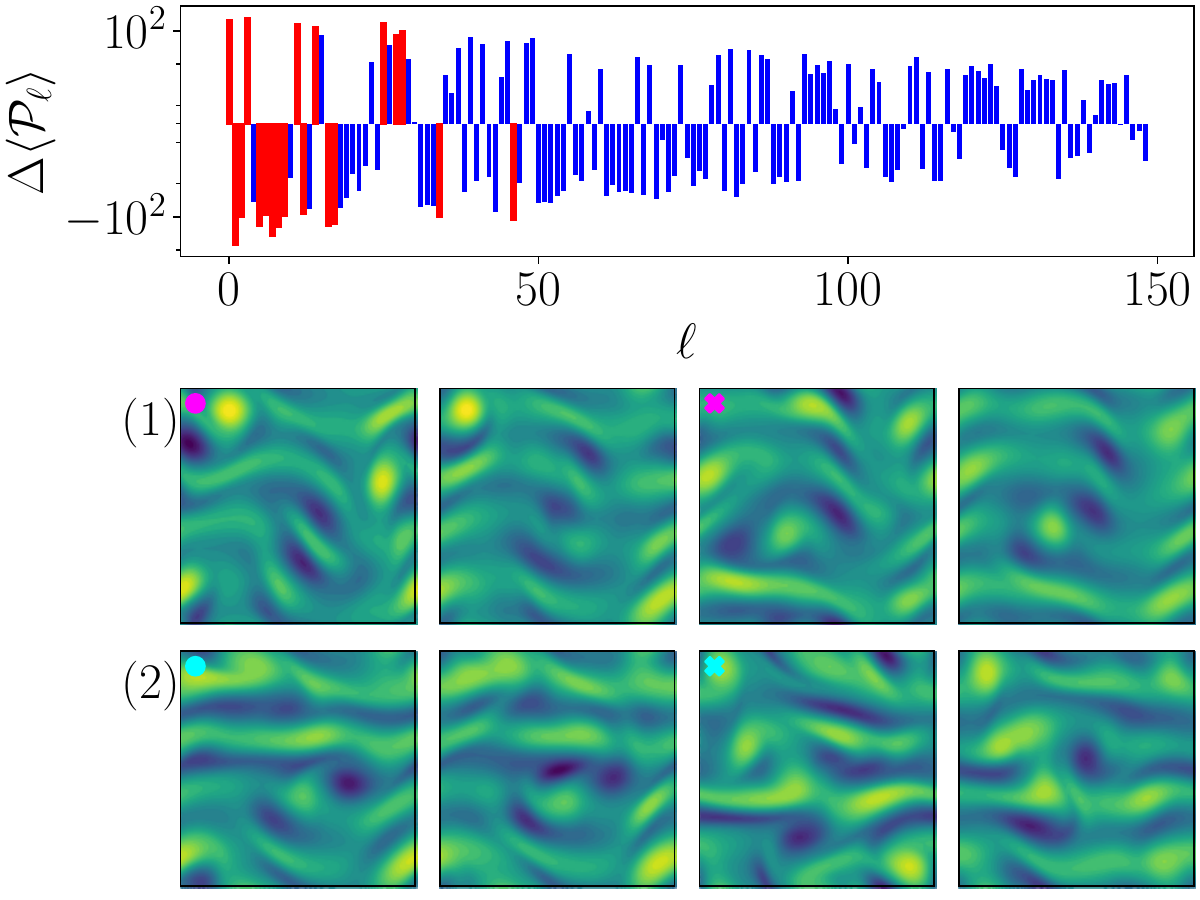}
    \end{minipage}
    \caption{Two-dimensional visualization of the co-ordinate vector (\ref{eq:tsne_in}) at $Re = 40$ using the t-SNE algorithm \citep{vandermaaten08}. (Left) The $x$- and $y$-axes represent the two degrees of freedom in the t-SNE visualization, which are determined such that the similarity between neighboring mapped data (measured by t-Student distributions) approximates the similarity in the original variables (measured by Gaussian distributions). Each snapshot is colored by a binned mean normalized dissipation rate $D / D_{lam}$ where $D_{lam} = Re/(2n^2)$ is the laminar dissipation rate, using 25 bins in both $x$ and $y$-directions. The dotted black lines indicate level curves of the Gaussian kernel density estimates of the representation, such that increasingly bold lines indicate increasing local density. The spanwise vorticity of representative states indicated by the \AC{arrows} visualize the distinct dynamical processes contained within the high density clusters of the visualization.
    Snapshots are extracted belonging to the bursting event class $i = 1,2$, and the average projection onto each mode $\langle \mathcal P_{\ell} \rangle_i$ of class $i$ is computed. (Right top) The difference $\Delta \langle \mathcal P_{\ell} \rangle \coloneq \langle \mathcal P_{\ell}\rangle_1 -\langle \mathcal P_{\ell} \rangle_2$ is shown for each mode $\ell$. A subset of modes $\ell \in L_d$ (red bars) appears to be mostly responsible for delineating the two bursting event classes.
    (Right bottom) Two representative snapshots $\omega$ indicated by fuchsia (top panels) and cyan (bottom panels) markers from the two classes of bursting events are shown. The decodes $\mathcal{D}(\sum_{\ell \in L_d} \mathcal{P}_{\ell}(\omega)u_{\ell} )$ are shown directly to the right of each representative snapshot $\omega$.
    }
    \label{fig:tsne_Re40}
\end{figure*}

The sparseness of the representation in the latent SVD basis motivates a lower-dimensional visualization. 
% now discuss the t-SNE visualization
We present two-dimensional representations of the modal projections of the embeddings (\ref{eq:tsne_in}) for each snapshot in the test dataset using the t-SNE algorithm \citep{vandermaaten08} for $Re = 40$, $Re = 100$ and $Re=400$.
\AC{As local pairwise distances between snapshots in the (truncated) latent space are `preserved' in the t-SNE visualizations \citep{vandermaaten08}, nearby snapshots in these visualizations reflect nearby snapshots in the truncated latent space.}
When constructing these visualizations, we also include previously computed libraries of unstable periodic orbits \citep[for full details see][]{page2022recurrent, Page2024, cleary2025} and 10 short trajectories of 100 advective time units sampled at intervals of 0.1 time units, when generating the t-SNE visualization.
These snapshots had to be included when running the t-SNE algorithm as it is a non-parametric self-supervised algorithm, but the UPOs and trajectory representations are only plotted in the low-dimensional visualizations when explicitly stated in the text. % (they are not shown in figures \ref{fig:tsne_Re40} and \ref{fig:tsne_Re100_400})}. 
The two-dimensional visualization plane is divided into $25\times25$ bins, and the mean dissipation rate of of the snapshots in each bin is computed.
Each point in the representation represents a snapshot in the dataset, and is colored by this local mean dissipation rate. 
The contours in figures \ref{fig:tsne_Re40} and \ref{fig:tsne_Re100_400} indicate level curves of the local density of the visualization, computed via Gaussian kernels \citep{scott2015}. 
The qualitative features (the clusters) of the output are robust to different initializations of the algorithm.

% discussion about the clustering in Re 40
The two-dimensional visualizations in figures \ref{fig:tsne_Re40} and \ref{fig:tsne_Re100_400} should be compared to those computed previously in PHBK and \cite{Page2021} -- these were dominated by the learned shift-reflect symmetry, and the data appeared to form a large octagon with little other structure.
The full symmetry reduction here, combined with the IRMAE approach, results in many more t-SNE clusters with distinct embedding features, each of which represents a distinctive `class' of vorticity snapshots in physical space. 

% [[still to edit the rest of this section]]}
At $Re = 40$, representative flow states are \AC{visualized in figure \ref{fig:tsne_Re40}} from a selection of low-dissipation t-SNE clusters, \AC{with their respective positions in the t-SNE visualization indicated by arrows.}
% Different combinations of physical features such as co-rotating vortex pairs, dominant vortex cores and diagonally stretched vortex cores are evident in each of the representative samples.
The clusters all show the characteristic zig-zag ``staircase'' patterns in the vorticity, and are distinguished by the number and relative locations of dominant vortices.
% Nearby clusters share some similar dynamical features indicating that the t-SNE algorithm has also managed to encapsulate some global structure. 
%The detached clusters represented by the snapshots are notably apart from the bulk of the visualization.
\AC{Two clusters on the left of the t-SNE visualization are notably detached from the bulk of the visualization.}
They sit far from the higher dissipation events and contain the lowest dissipation fields with large patches of quiescent fluid \AC{(see the representative snapshots with a white circle in the top left corner)}.
% In contrast, the clusters indicated by the brown pentagon and yellow right-pointing triangle are much closer to the bursting states, suggesting that states within these clusters are possibly more prone to bursting. 
The clear clustering into many distinct classes of events -- not visible in previous work \cite{Page2021,Page2024} -- and apparent distinction of different types of high-dissipation dynamics
% This global structure of the visualization 
motivates further investigation into the nature of bursting events and routes into/out of this region of state space.

\subsubsection{\AC{Bursting classes}}

% bursting events discussion
The high-dissipation \AC{or `bursting'} events (yellow/orange/red markers in figure \ref{fig:tsne_Re40}) are found in two (possibly three) distinct \AC{classes}: one surrounded by lower-dissipation events, labeled `1' in figure \ref{fig:tsne_Re40}, and another which sits apart in the visualization and is more diffuse, labeled `2'.
% Firstly, there appear to be two distinct classes of high-dissipation snapshots at $Re=40$, separated by a small `band' of lower-dissipation snapshots -- a distinction which was not apparent in earlier models \cite{Page2021,Page2024}}.
% The smaller class within the bulk of the visualization is a single cluster of moderately high dissipation states, while the larger class appears to be comprized of two detached clusters of stronger bursts. 
This second, higher-dissipation region may itself be made up of two sub-classes, though the amount of data in this region makes this difficult to discern.
The \AC{two classes of bursting events can be distinguished by a} subset of latent SVD modes $\ell \in L_d$.
This is done in \AC{the right panel of} figure \ref{fig:tsne_Re40}, 
% , and the physical structures encoded by these modes are notably different for the two classes. 
where the most significant modes are identified by computing the mean projection onto each $\ell$ within the class, $\langle \mathcal P_{\ell}(\omega) \rangle_i$ ($i=1$ or 2) (the average was taken over all snapshots satisfying $D/D_{lam}> D_c$ within manually defined boundaries on the t-SNE plane).
The subset of modes with the largest difference $\Delta \langle \mathcal P_{\ell} \rangle \coloneq \langle \mathcal P_{\ell}\rangle_1 -\langle \mathcal P_{\ell}\rangle_2$ are highlighted in red in \AC{the right panel of} figure \ref{fig:tsne_Re40} -- this defines the subset $L_d$.

% by the mean projection onto each mode $\langle \phi \rangle_i$ for both classes of bursting events $i = 1,2$. 
% The mean projections $\langle \phi \rangle_i$ were computed by averaging over all snapshots satisfying $D/D_{lam}> D_c$ within manually defined boundaries of each bursting class.}
% The subset of modes with the largest difference $\Delta \langle \phi \rangle \coloneq \langle \phi\rangle_1 -\langle \phi\rangle_2$ are then extracted, and highlighted in red in the top panel of figure \ref{fig:tsne_bursting}.
% %smaller high-dissipation cluster (marked by the fuchsia cross in figure \ref{fig:tsne_bursting}) has a much larger projection onto this mode than the larger high-dissipation cluster (cyan cross).

The physical structures spanned by these modes are revealed in \AC{the bottom right panel of} figure \ref{fig:tsne_Re40}, by visualizing the decode of the sum of the projection onto these modes $\mathcal{D}(\sum_{\ell \in L_d} \mathcal{P}_{\ell}(\omega_i)u_{\ell} )$ for representative snapshots $\omega_i$ from the two bursting classes which \AC{are indicated in the t-SNE plot.}
Class 1 (the calmer high-dissipation cluster) has a stronger contribution from the SVD mode 0, which was associated with the time-averaged mean flow (see figure \ref{fig:modes}). 
% AC: average projections for both class 1 and 2 are positive for first three modes. this makes it okay to say "stronger" projection. 
%and staggered diagonal stripes of vorticity 
The signature of this large-scale mode is clearly visible in the decodes of the subset $L_d$ in {the bottom right panels of} figure \ref{fig:tsne_Re40}. 
% These decodes suggest that snapshots belonging to the calmer cluster of high dissipation states are quite closely related to the mean flow state at $Re = 40$.
In contrast, there is a weaker projection onto this mode for class 2, the higher dissipation class, which explains its location away from the bulk of the visualization in figure \ref{fig:tsne_Re40}.
Class 2 instead has a stronger projection onto more of the modes $\ell \in L_d$, and also excites more large $\ell$ modes which describe smaller-scale vorticity features.

\subsubsection{\AC{Bursting trajectories}}

\begin{figure}%[!ht]
    \centering
    \includegraphics[width=\linewidth]{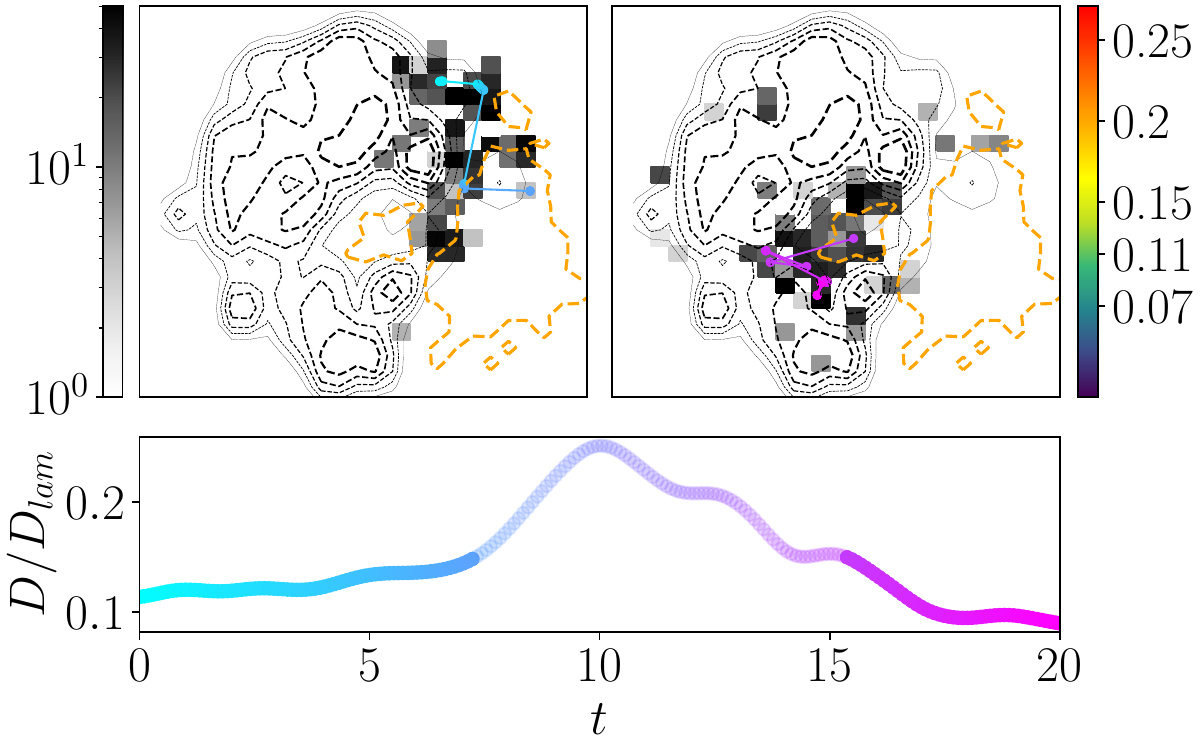}
    \includegraphics[width=\linewidth]{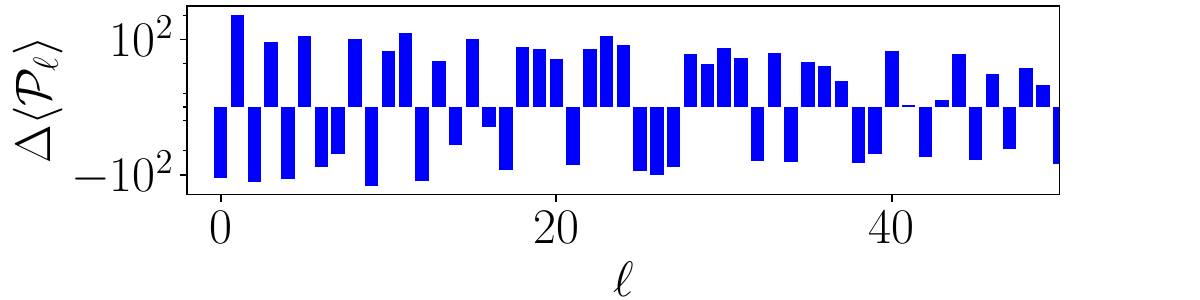}
    \caption{(Top) Two-dimensional visualizations of the co-ordinate vector (\ref{eq:tsne_in}) at $Re = 40$ using the t-SNE algorithm \citep{vandermaaten08}. The dotted black lines indicate level curves of the Gaussian kernel density estimates of the representation, with increasingly bold lines indicating increasing local density. The dotted orange line identifies the level curve of $D / D_{lam} = 0.15$.
    The grayscale histogram indicates the distribution of subtrajectories \AC{as defined in the text} (left) which precede and (right) follow a bursting event. 
    %, which is defined by a local maximum in the dissipation rate satisfying $D > 0.175D_{lam}$. 
    %Each subtrajectory is defined by the closest continuous sequence of snapshots satisfying $D < 0.15D_{lam}$ to the local maximum, within 10 advective time units. The subtrajectory is also required to start from or finish at a sufficiently quiescent state satisfying $D < 0.125D_{lam} $. %length 10 advective time units directly (left) before and (right) after a bursting event, which is defined by a local maximum in the dissipation rate. 
    (Middle) The dissipation of a sample subtrajectory centered on the bursting event is plotted against time in advective time units, colored according to time. The opaque markers denote the pre- and post-bursting subtrajectories extracted, while the transparent markers indicate the neglected bursting event. The same subtrajectories are shown by the connected, similarly colored points in the t-SNE visualization, demonstrating these sample routes (top left) towards and (top right) away from the bursting event.
    (Bottom) The difference $\Delta \langle \mathcal P_{\ell} \rangle \coloneq \langle \mathcal P_{\ell}\rangle_{b} - \langle \mathcal P_{\ell} \rangle_{turb}$ is shown for each mode $\ell$, where $\langle \mathcal P_{\ell}\rangle_{b}$ and $\langle \mathcal P_{\ell} \rangle_{turb}$ denote the mean projection of snapshots belonging to the pre-bursting cluster (red \AC{star} in figure \ref{fig:tsne_Re40}) and all snapshots in the test dataset, respectively.
    %(Bottom) Each row of snapshots corresponds to equally spaced snapshots sampled along a representative subtrajectory directly (first row, pink markers) before and (second row, cyan markers) after the local maximum in dissipation. The snapshots are sampled $2$ advective time units apart along the subtrajectory, denoted by the plus, cross and diamond markers, with dissipation rates (first row, pink markers) $D/D_{lam} = 0.094, 0.113, 0.142$ and (second row, cyan markers) $D/D_{lam} = 0.228, 0.171, 0.123$ respectively. 
    \label{fig:tsne_trajs}}
\end{figure}

% routes into bursting
Plotting highly temporally-resolved trajectories in the t-SNE visualizations can be used to identify routes into and out of the high dissipation regions, i.e.\ `bursting' events. 
This allows us to probe whether either bursting class is visited in isolation, and where likely bursts originate from in state space.
% Secondly, the transition routes between low and high dissipation events at $Re = 40$ are investigated. 
Within the t-SNE dataset are 10 trajectories of length 100 advective units, and from these we extract and visualize all of the subtrajectories which either lead to bursting events from quiescent events, or vice-versa.
These subtrajectories are defined as the continuous sequences of snapshots satisfying $D < D_{c}$ which precede or follow the local dissipation maximum, where $D_c$ is the same dissipation rate threshold used for the dataset resampling in \S\ref{sec:flow_conf}.
We also require that the subtrajectories start from or finish at a sufficiently quiescent state satisfying $D < D_{l}$ and that the local dissipation maximum satisfies $D > D_h$, where $D_l, D_h = 0.125D_{lam}, 0.175D_{lam}$ for $Re= 40$.
% By separating the dynamics directly before a bursting event from the dynamics directly after a bursting event, we can identify the routes into and out of the high dissipation region on the two-dimensional representation generated by the t-SNE algorithm in figure \ref{fig:tsne_trajs}.
The dissipation rate and low-dimensional representation of one such bursting subtrajectory is shown in the middle panel of figure \ref{fig:tsne_trajs}, where the dissipation rates of the pre-bursting and post-bursting snapshots are indicated by the opaque markers. 

% At $Re = 100$, the visualization cannot delineate between regions of bursting events from either pre-bursting or post-bursting events.
% There are likely many different physical routes to the wide range of bursting dynamics at this $Re$, such that this low-dimensional visualization struggles to separate these dynamics. 
Histograms of all pre- and post-bursting events, visualized over the earlier t-SNE map of the state space, are reported in \AC{the top left and right panels of} figure \ref{fig:tsne_trajs}, \AC{respectively}.
% The t-SNE visualization clearly distinguishes the dynamics leading to a bursting event at $Re = 40$ from the dynamics after a bursting event in figure \ref{fig:tsne_trajs}.
There is a clear distinction between pre- and post-bursting events.
The plot reveals that that the `calmer' bursting class \AC{1} is not visited in isolation, but is the route of all high-dissipation events back to quiescent dynamics.
% Bursting events do not visit the calmer bursting class in isolation, but instead visit this class as they settle back to quiescent behavior. 
This is confirmed by inspecting the full trajectories of bursting events on the t-SNE visualization (not shown), which always first hop between the two clusters within the extreme bursting class 2, before moving to the calmer bursting class 1.

The pre-bursting trajectories always visit the t-SNE cluster marked by a red \AC{star} in figure \ref{fig:tsne_Re40}, indicating that this cluster encapsulates some important pre-bursting vorticity features. 
To identify the associated distinguishing latent modes, the mean projection onto each mode $\ell$ for the snapshots within this cluster $\langle \mathcal P_{\ell} (\omega)\rangle_b$ and for all the snapshots in the test dataset sampled from the turbulent attractor $\langle \mathcal P_{\ell} (\omega)\rangle_{turb}$ are computed. The differences between these average projections are visualized in the lower panel of figure \ref{fig:tsne_trajs}.
This computation indicates that the pre-bursting cluster has a weaker projection onto the $\ell = 0$ mode representative of typical low-dissipation states.
However, a much stronger projection onto the $\ell = 1$ mode, which features  staggered diagonal stripes of vorticity (see figure \ref{fig:modes}) similar to unstable directions associated with many exact solutions which resemble the mean flow state (see figure \ref{fig:TW_stab}), is the distinguishing feature of the pre-bursting dynamical process within this cluster. 

%[[Andrew: can you connect back to the cluster labels in the earlier figure, e.g. is it the `yellow triangle' that is the pre-burst class? If so, we should make this connection and mention what the physical structure is, which latent modes it highlights etc. The Fourier modes -- while useful -- are almost an argument against our approach, we need to use the latent vectors to say something too, otherwise why bother.]}
We have also confirmed that pre-bursting and post-bursting regions on the visualization are both characterized by small $| \hat{\omega}(1,0)|$, while the bursting regions have a large value for $| \hat{\omega}(0,n) |$ (not shown).
This large value in the bursting region is to be expected, as the production rate of Kolmogorov flow can be expressed by $I = -\hat{\omega}_{R}(0,n) / n$. 
The small value of $| \hat{\omega}(1,0)|$ in the pre-bursting region and the large value of$| \hat{\omega}(0,n) |$ in the bursting region confirms the significant redistribution of energy between these modes before a bursting event, an observation which was also made in the variational approach of \citep{farazmand2017}.

\subsubsection{\AC{Higher $Re$ visualizations}}
\begin{figure}
    \centering
    \includegraphics[width=\linewidth]{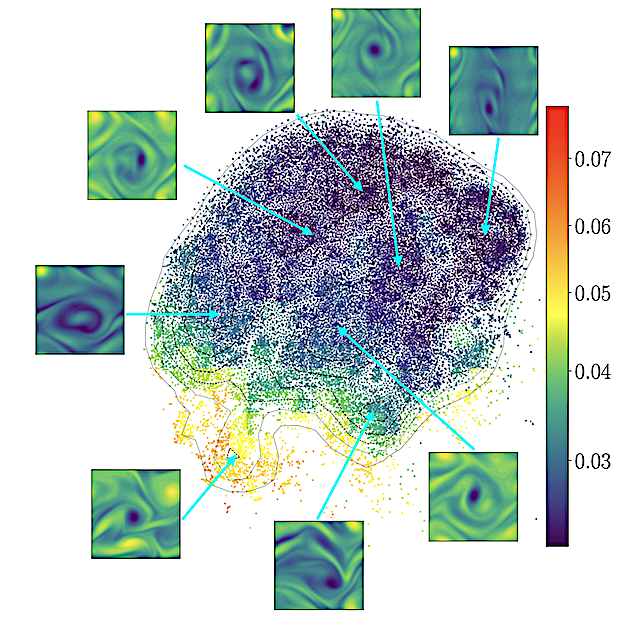}
    \rule{\linewidth}{0.5pt}
    \includegraphics[width=\linewidth, trim={0 0.2cm 0 -0.2cm }, clip]{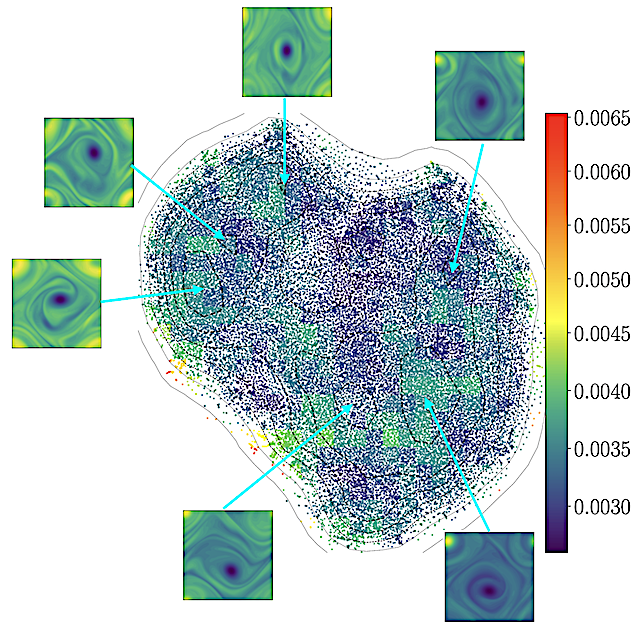}
    \caption{Two-dimensional visualizations of the co-ordinate vector (\ref{eq:tsne_in}) at (top) $Re = 100$ and (bottom) $Re = 400$ using the t-SNE algorithm \citep{vandermaaten08}, as in figure \ref{fig:tsne_Re40}\AC{, with visualizations of representative snapshots from the dense clusters.}
    \label{fig:tsne_Re100_400}}
\end{figure}

% Re 100 visualization
At $Re = 100$ (top panel of figure \ref{fig:tsne_Re100_400}), many high-dissipation snapshots are spaced across the lower periphery of the two-dimensional visualization. 
There is also a distinct detached high-dissipation cluster, indicating a distinct class of events which could not be separated cleanly in earlier models \cite{Page2021,Page2024}. 
% We note that the high dissipation events could not be separated cleanly from the low dissipation snapshots at $Re = 100$ without the full symmetry reduction \citep{Page2024}. 
% Even though the flow at $Re = 100$ is considerably more chaotic than at $Re = 40$, there are still clear high density clusters visible, representing distinct dynamical processes. 
Although the clustering of data is less obvious than at the lower $Re=40$ value, there are still distinct classes of vorticity snapshots that can be identified in this visualization.
Some representative snapshots sampled from these clusters are shown in \AC{the top panel of figure \ref{fig:tsne_Re100_400}}. 
% The top row of snapshots shows examples from the two larger clusters in the middle of the t-SNE visualization, which demonstrate a nice variety of dynamical processes.
% A high density cluster to the extreme right of the t-SNE visualization represents the very low dissipation rate events, and exhibits the alternating horizontal bands of vorticity closely related to the Kolmogorov forcing profile. 
% Two smaller high density regions closer to the high dissipation periphery of the visualization are also presented, which exhibit much more chaotic dynamics including a strong vorticity ring. 
% These events are likely dynamically linked to the wide range of bursting events observed at $Re = 100$. 
% A sample from the highest density cluster in the high dissipation periphery is also shown in the final panel of figure \ref{fig:tsne_100egs}. 
% The dynamics within this cluster is highly varied, but typically features at least two very strong counter-rotating vortices. 
This figure indicates that the classes can be delineated based on the appearance of the large scale vortices in the flow -- their number, relative location and strength.
The sample from the detached high-dissipation cluster \AC{(bottom left of the visualization)} again indicates an event where the importance of `mean flow' modes (e.g. $\ell = 0$ and 1) is diminished. 

% \begin{figure}%[!ht]
%     \centering
%     \includegraphics[width=\linewidth]{Figures/Re100_tsne_egs_90.pdf}
%     \caption{Representative snapshots from the dense clusters in the t-SNE visualization of the co-ordinate vector (\ref{eq:tsne_in}) at $Re = 100$. The position of each snapshot in the visualization is marked by the correspondingly colored marker in figure \ref{fig:tsne_Re100_400}.
%     \label{fig:tsne_100egs}}
% \end{figure}

% Re 400 visualization 

Finally, the t-SNE visualization of the most turbulent flow considered at $Re = 400$ (bottom panel of figure \ref{fig:tsne_Re100_400}) no longer appears to cluster events based on a dominant dissipation rate. 
The high dissipation snapshots are distributed fairly uniformly in the latent t-SNE visualization.
% The binned dissipation rate is quite low and smooth across the visualization because the high dissipation snapshots are distributed throughout the visualization, such that there are no localized regions of high dissipation bursts. 
The exception here are a couple of small pockets of very high dissipation snapshots at the left periphery of the visualization. 
% The relatively large reconstruction error of this model is likely reflected in this poor latent representation of the snapshots. 
% More sophisticated network architectures or larger training datasets are likely required to improve the reconstruction error, before any further analysis is performed at this very high $Re$.
This behavior is consistent with the prominence of the large vortex pair associated with the condensate, a state which is apparent in most snapshots and hence leads to a more homogeneous latent representation across the dataset.
Some representative snapshots of the high density regions of the t-SNE visualization are \AC{again presented in the bottom panel of figure \ref{fig:tsne_Re100_400}}, and all show a large scale vortex pair.
Distinctions between snapshots are likely based on smaller scale features (e.g. vortex filaments) and the relative location/shape of the dominant vortices.

% \begin{figure}%[!ht]
%     \centering
%     \includegraphics[width=\linewidth]{Figures/Re400_tsne_egs_90.pdf}
%     \caption{Representative snapshots from the dense clusters in the t-SNE visualization of the co-ordinate vector (\ref{eq:tsne_in}) at $Re = 400$. The position of each snapshot in the visualization is marked by the correspondingly colored marker in figure \ref{fig:tsne_Re100_400}.
%     \label{fig:tsne_400egs}}
% \end{figure}

\subsection{Periodic orbit coverage}
% periodic orbits discussion
\begin{figure}%[!ht]
    \centering
    \includegraphics[width=0.5\linewidth]{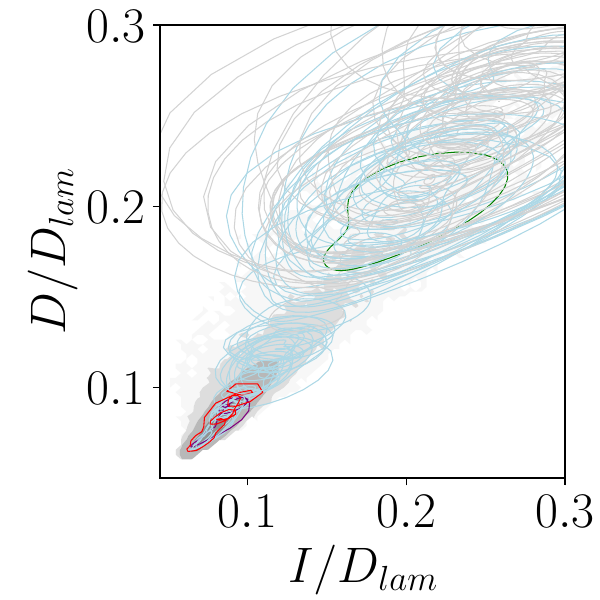}%
    \includegraphics[width=0.5\linewidth]{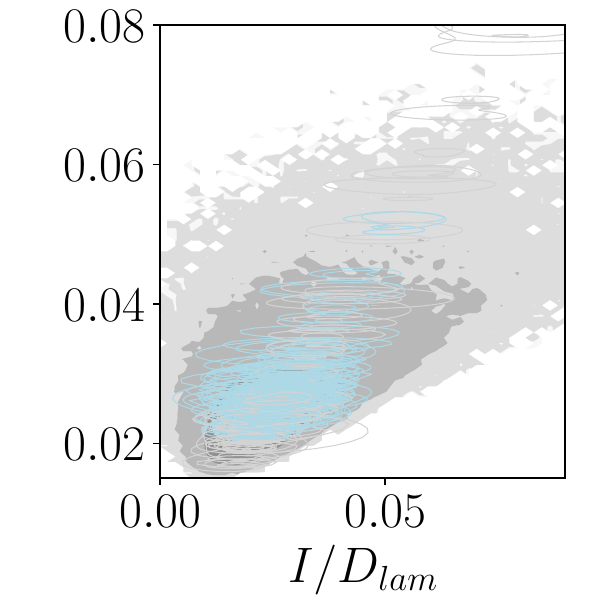}
    \caption{Dissipation rate $D$ against production rate $I$ of the UPOs in the libraries of (left) 174 solutions at $Re = 40$ and (right) 252 solutions at $Re = 100$, both normalized by the laminar dissipation $D_{lam} = \Rey / (2n^2)$. 
    The gray background is the probability density function (PDF) computed from a trajectory of $10^5$ samples, separated by 1 advective time unit. The contour levels of the PDF are spaced logarithmically. Solutions deemed `dynamically relevant' by the simple dissipation/energy comparison described in the text are colored in light blue, while those note satisfying the condition are colored in gray.}
    \label{fig:diss_prod}
\end{figure}

The motion of a turbulent flow within the turbulent attractor is postulated to be shaped and guided by unstable periodic orbits (UPOs) -- exact solutions of the Navier-Stokes equations that each encapsulate some closed cycle of physical processes relevant to the turbulence itself \cite{cvitanovic2010,Kawahara2012,Graham2021}. 
A key motivation for this viewpoint is the possibility to connect dynamical events in the flow to statistical properties of turbulence, e.g. the cascade(s), via periodic orbit theory \cite{Artuso1990a,Artuso_1990b} which provides weights for expansions of the statistics of a chaotic attractor in terms of the statistics of the UPOs.

Large numbers of solutions have been amassed recently for the Kolmogorov flow studied here \cite{page2022recurrent,Page2024, cleary2025} at $Re \in \{40, 100\}$. 
These solutions are visualized in figure \ref{fig:diss_prod} in terms of their production/dissipation rates, overlayed on the turbulent distribution. 
These low dimensional visualizations, and others like them, would seem to indicate the `dynamical relevance' \cite{Krygier2021} of the solutions due to their apparent proximity to the turbulence. 
However, these visualizations are not nuanced enough to identify the range of events contained in the library of UPOs, and may give a misleading impression as to the relevance of a given solution to the true turbulent dynamics. 
The t-SNE plots constructed above provide a way to \AC{visualize} the degree to which the range of true vortical events in the flow is covered by the library of solutions. 
\AC{As nearby snapshots in the t-SNE visualizations reflect similar snapshots in the truncated latent space, the visualizations effectively reveal whether the UPOs are similar to samples from the chaotic attractor and provide an alternative way to assess dynamical relevance.}
% \AC{This dynamical relevance condition is based on the vortical structures in instantaneous snapshots sampled along the periods of the UPOs and sampled from the chaotic attractor.}

%In recent years, there has been a large push to compute unstable periodic orbits (UPOs) in various flows, such as parallel pipes \citep{Kerswell2005,Budanur2017}, channels \citep{Waleffe2001,Park_Graham_2015}, plane Couette flow \citep{Gibson2008, Krygier2021}, 
%%rotating superfluids \citep{campbell1979vortex, Cleary2023}, 
%periodic, body-forced two-dimensional \citep{Chandler2013, lucas2015, suri2020, page2022recurrent, Zhigunov2023, Redfern2024, cleary2025} and three-dimensional flows \citep{yalniz2021, lucas2017}, among many others. 
%This has been motivated in part by the possibility of predicting turbulent statistics, such as energy distributions or velocity moments, using a weighted sum of these UPOs. 
%Periodic orbit theory formalizes the computation of these weights in uniformly hyperbolic dynamical systems \citep{ChaosBook, Artuso1990a, Artuso_1990b, CVITANOVIC1991}.
%Previously, this approach had been limited by a paucity of solutions, until a recent work \citep{page2022recurrent} proposed a robust approach to compute large numbers of exact solutions using gradient-based optimization.
%This approach was also applied to compute very large numbers of exact solutions of rotating superfluids \citep{cleary2025}.
%Here, we investigate the dynamical relevance of previously computed solutions at $Re = 40$ and $Re = 100$. 

\subsubsection{\AC{$Re = 40$ UPOs}}

\begin{figure}%[!ht]
    \centering
    \includegraphics[width=\linewidth]{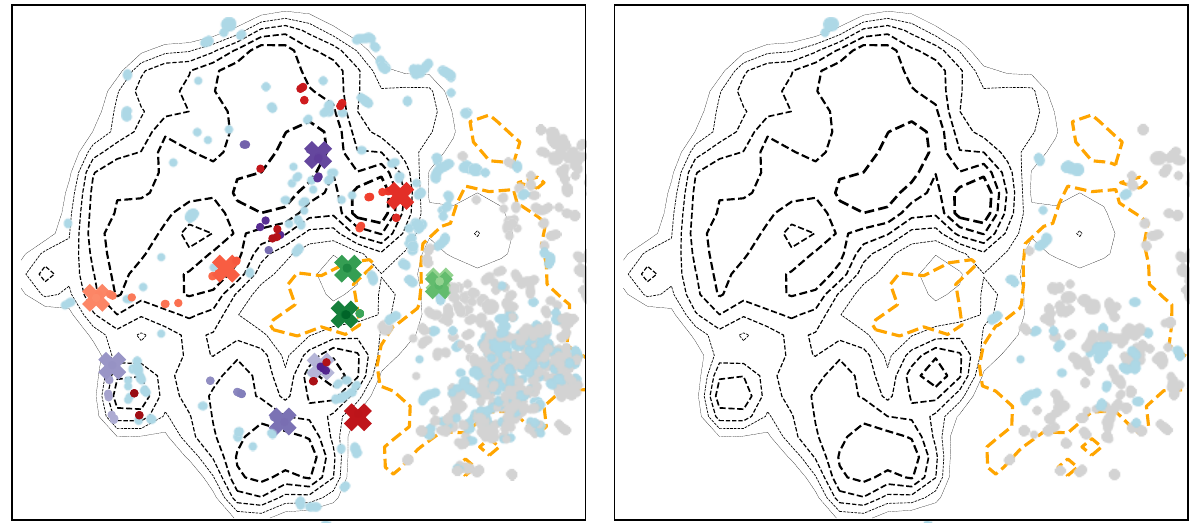}
    \includegraphics[width=\linewidth]{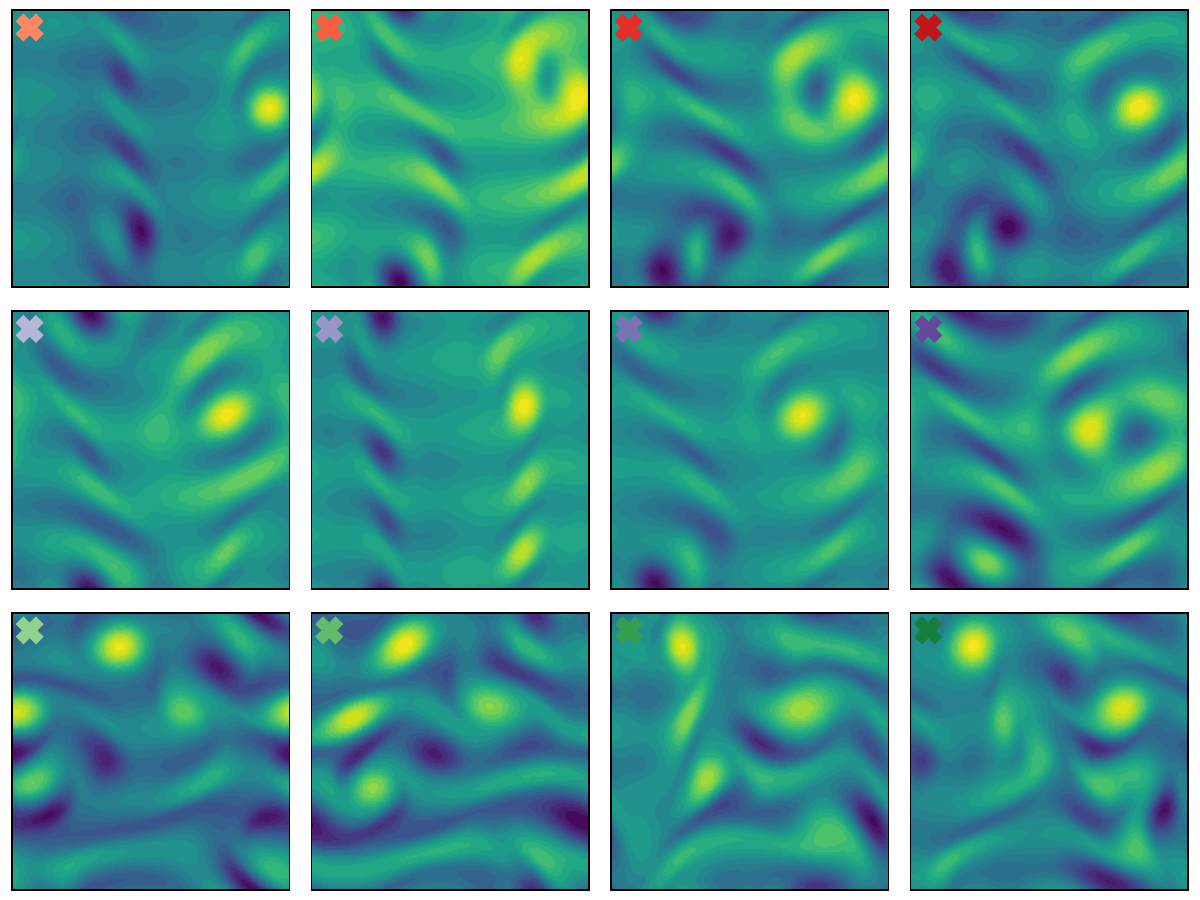}
    \caption{(Top) Two-dimensional visualizations of the co-ordinate vector (\ref{eq:tsne_in}) at $Re = 40$ using the t-SNE algorithm \citep{vandermaaten08}. The dotted black lines indicate level curves of the Gaussian kernel density estimates of the representation, such that increasingly bold lines indicate increasing local density. The dotted orange line indicates the level curve of $D / D_{lam} = 0.15$. The points indicate snapshots sampled along the periods of the `dynamically relevant' (light blue) and `dynamically irrelevant' (gray) UPOs in (left) the full library of 174 solutions and (right) the short-period orbits with periods $T_{upo} < 5$ advective time units. 50 snapshots are sampled for each UPO. 
    (Bottom) Each row shows equally spaced snapshots (without full symmetry preprocessing) along the periods of three periodic orbits with periods $T_{upo} = 23.16, 18.77$ and $5.06$, indicated in the t-SNE visualization by the red, purple and green points respectively, from top to bottom, where increasing boldness denotes time evolution along the period of the orbit. The crosses in the right top panel indicate the positions of the corresponding visualized snapshots. 
    \label{fig:tsne_upos}}
\end{figure}

% separate Re = 40 discussion from Re = 100 
The $Re=40$ UPOs are visualized within the two-dimensional t-SNE visualization in figure \ref{fig:tsne_upos}. 
The solutions are colored gray or blue according to a simple a priori `dynamical relevance' criteria
of whether the time-averaged dissipation rate $\langle D \rangle$ and energy $\langle E \rangle$ lie within the $1^{st}$ and $99^{th}$ quantiles of the corresponding turbulent distributions.
While some high-density regions of the turbulence feature snapshots from the UPOs, there are large areas of high turbulent-snapshot density which do not contain any at all.
% I would say really that coverage is quite poor everywhere -- I'll comment this out and add a line in the following paragraph about some high dissipation states potentially being relevant in t-SNE but not in I-D
% In particular, there is very poor coverage of the high-dissipation t-SNE clusters visible in figure \ref{fig:tsne_Re40} -- the UPOs are mostly confined to the sparse region of infrequent bursting states.}
This is surprising given the robustness of the statistical reconstructions possible with this set of solutions observed in \cite{page2022recurrent,cleary2025}.

When only the shorter orbits (here $T < 5$) are visualized (see top right panel of figure \ref{fig:tsne_upos}) the picture simplifies further.
%The short period orbits tend to be higher dissipation and only a small number appear to lie close to the high dissipation t-SNE clusters.} 
The low-dissipation, short orbits are confined to the edges of the visualization.
On the other hand, there are many high-dissipation UPOs which were previously labeled `irrelevant' (based on $E$ and $D$ -- gray symbols) that do overlap with the t-SNE embeddings of turbulent snapshots. 
This indicates that turbulent orbits may visit the neighborhoods of these solutions despite the mismatch in their production/dissipation values. 
%[[JP: Actually Andrew, there is a problem here: if I look at the original t-SNE then I do see many high dissipation snapshots outside of the kernel density estimation. So I don't think it's fair to say these events are dynamically irrelevant. In fact, it would seem that some of the solutions which are high dissipation and which the naive metric says are not dynamically relevant ARE in fact close to turbulence.]]}
Snapshots from some individual -- modest length -- UPOs are also included in figure \ref{fig:tsne_upos} and are identified in the t-SNE visualization.
These UPOs are also colored accordingly in the dissipation versus production rate diagram in the left panel of figure \ref{fig:diss_prod}.
The two longest period orbits, with period $T = 23.16$ (red) and $T = 18.77$ (purple) visit many of the high density clusters throughout the visualization. 
In contrast, the $T = 5.06$ (green) orbit is shorter and is more localized to a high dissipation region of the visualization.
This UPO is the only solution in the library which visits the smaller bursting cluster at $Re = 40$ discussed in section \ref{sec:low-d-v}.

\subsubsection{$Re = 100$ UPOs} 

\begin{figure}%[!ht]
    \centering
    \adjincludegraphics[width=\linewidth, trim={0 0 {.5\width} 0},clip]{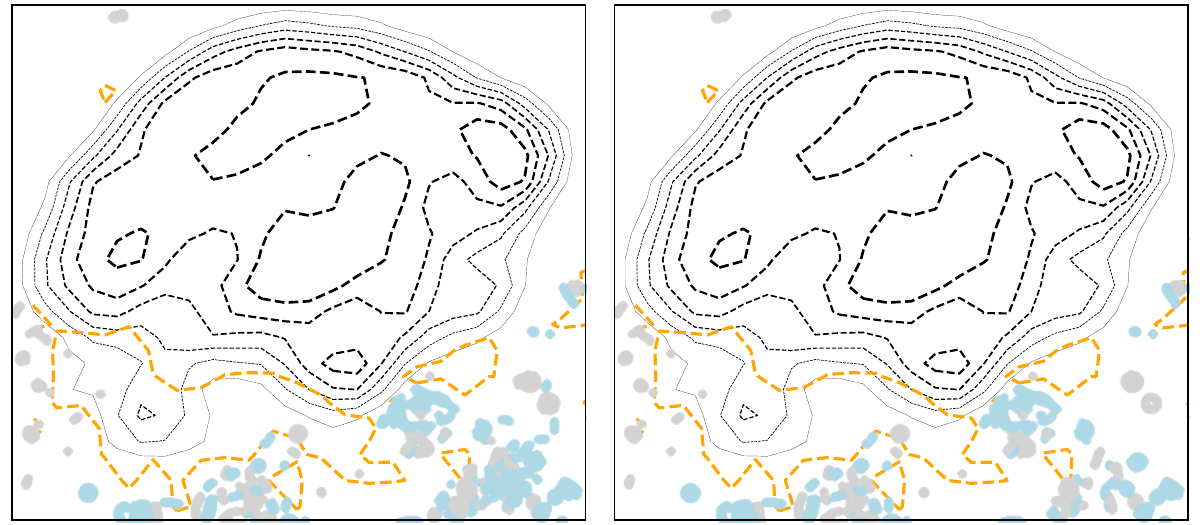}
    \includegraphics[width=\linewidth,trim={0 0 0.425cm 0},clip]{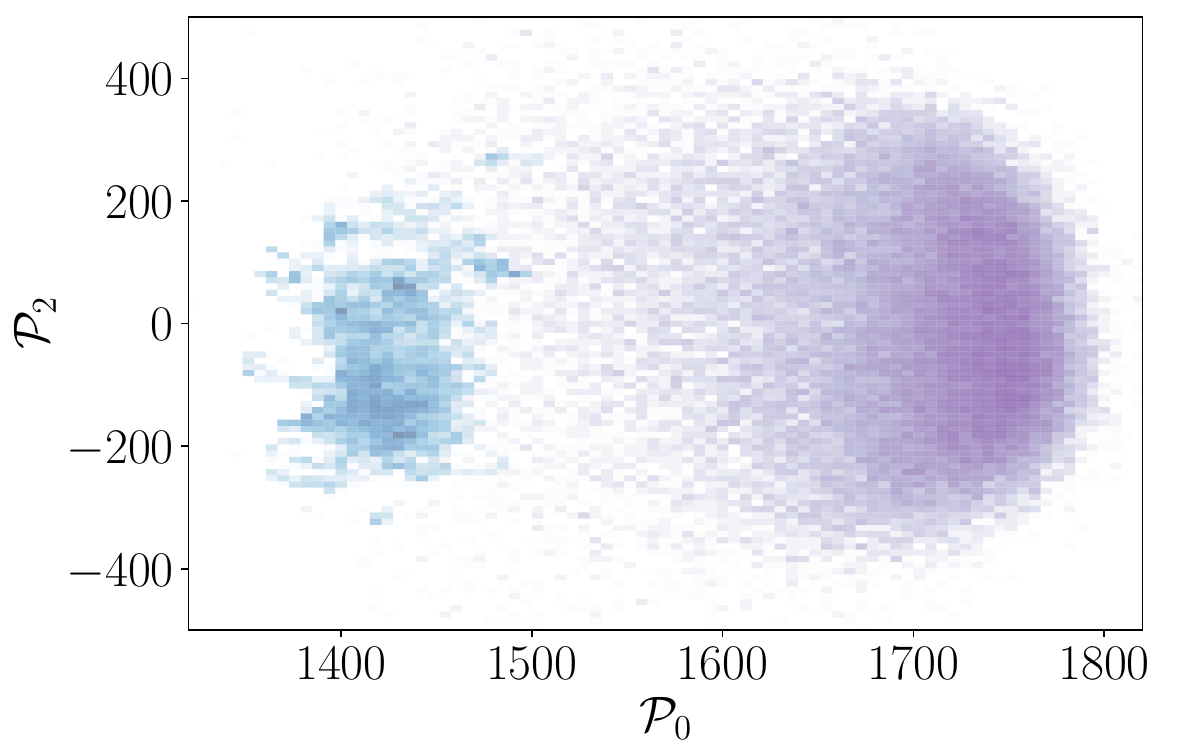}
    \caption{(Top) Two-dimensional visualizations of the co-ordinate vector (\ref{eq:tsne_in}) at $Re = 100$ using the t-SNE algorithm \citep{vandermaaten08}. The dotted black lines indicate level curves of the Gaussian kernel density estimates of the representation, such that increasingly bold lines indicate increasing local density. The dotted orange line indicates the level curve of $D / D_{lam} = 0.045$. The points indicate snapshots sampled along the periods of the `dynamically relevant' (light blue) and `dynamically irrelevant' (gray) UPOs in the full library of 252 solutions. 50 snapshots are sampled for each UPO. (Bottom) The joint distribution of the latent projections $\mathcal{P}_0$ and $\mathcal{P}_2$ of snapshots sampled from the periods of the `dynamically relevant' UPOs (blue) and snapshots sampled from the turbulent attractor (purple). The densities are colored on a log scale.
    \label{fig:Re100_tsne_upos}}
\end{figure}

A similar visualization is conducted for the UPOs computed at $Re=100$ in \AC{the top panel of figure \ref{fig:Re100_tsne_upos}}.
%Perhaps even more strikingly at $Re = 100$ in figure \ref{fig:Re100_tsne_upos}, nearly all the UPOs computed in \citep{page2022recurrent, cleary2025} lie outside the lowest density contour. 
%The vast majority of these solutions are short-period orbits. 
While the vast majority of UPOs have production/dissipation values consistent with that of the turbulence (see figure \ref{fig:diss_prod}), the t-SNE map indicates that nearly \emph{all} UPOs are determined by the autoencoder to be distinct from the turbulent dataset. 
% not sure we should make the statement below -- we know we are missing lots of short ones from Jeremy 
% It should be noted that the majority of these UPOs are short period orbits ($T < 5$), again hinting at an apparent dynamical irrelevance of short period orbits of Kolmogorov flow.}
This is quite striking given the overlap in various diagnostic statistics reported here and elsewhere \cite{Chandler2013,page2022recurrent,cleary2025}.
The apparent lack of similarity can be diagnosed by considering the projections $\mathcal P_{\ell}(\omega)$ onto individual latent directions.

To investigate the cause for the separation of these UPOs from the turbulent attractor, the mean projection onto each mode $\ell$ was computed for all the snapshots sampled from the `dynamically relevant' (as determined naively from the dissipation and energy PDFs) UPOs at $Re = 100$, $\langle \mathcal P_{\ell}(\omega) \rangle_{UPO}$, and for all snapshots in the test dataset sampled from the turbulent attractor at $Re = 100$, $\langle \mathcal P_{\ell}(\omega) \rangle_{turb}$.
% [[Andrew -- do you mean $\phi$ or $\mathcal P_l$?]]}
% [[We compute the projection onto every non-negligible mode $\ell < c$, which is why I wrote $\langle \phi \rangle$ here,  rather than using $\mathcal P_l$ which might imply \emph{all} modes $\ell$. But perhaps $\mathcal P_l$ works better in this discussion as we don't show the differences in $\langle \phi \rangle$, as we do in previous sections?]]}
%This analysis, similar to the modal projection analysis in section \ref{sec:low-d-v}, revealed the most distinct modes between $\langle \phi \rangle_{UPO}$ and $\langle \phi \rangle_{turb}$.
Differences were found in the distributions for large numbers of the latent projections, but the most significant distinction between the turbulence and the UPOs was found in the leading mode $\ell = 0$.
Joint distributions for both the UPO data and turbulence are reported in \AC{the bottom panel of figure \ref{fig:Re100_tsne_upos}} (the other mode is selected as $\ell = 2$ here).
This figure highlights the substantial difference in the $\ell = 0$ distributions -- the UPO/turbulence data overlap only slightly in the tails of the distributions. 
Recall that in \S \ref{sec:latent_structure} the $\ell = 0$ mode was decoded and shown to be visually similar to the symmetry-reduced, time-average flow state in the turbulence. 
The dramatic reduction of importance of this mode should also be considered in the context of the continuation study reported in \cite{cleary2025}, which showed that many $Re=100$ solutions appear to `move away' from the turbulent attractor as $Re$ increases. 
The results reported here would suggest that perhaps they are not inside the turbulent attractor, but just `close by' at $Re=100$ -- consistent with the continuation results.  

\section{Conclusion}

In this study we designed deep autoencoders for learning low-dimensional representations of 2D Kolmogorov flow. 
The models combined (i) full discrete/continuous symmetry reduction, (ii) dense convolutional input/output streams and (iii) an `implicit rank minimizing' (IRMAE \cite{jing2020,zeng2021}) component.
The autoencoders were used to estimate the dimensionality of the turbulent attractor in the two-dimensional Navier-Stokes equations -- which was obtained from the rank of the inner-most network representation.
% This study presented the estimation of an upper bound to the dimension of the invariant manifold $\Tilde{d}_{\mathcal{M}}$ of two-dimensional $n = 4$ Kolmogorov flow using convolutional autoencoders with implicit rank minimization.
An approximate scaling of $\Tilde{d}_{\mathcal{A}} \sim Re^{1/3}$ was found in the fully turbulent regime -- a much weaker scaling than the known upper/lower bounds on the \emph{global} attractor \citep{CONSTANTIN1988, DOERING1991, Liu1993}.
% A variety of techniques were introduced to improve the reconstruction accuracy and enstrophy spectrum of the trained autoencoders, including training data resampling, full symmetry reduction and a stream of output layers of decreasing convolutional filter sizes. 

The combination of the IRMAE approach with a full symmetry reduction yielded a rich latent-space structure, with which individual latent SVD modes could be decoded into physically interpretable structures at all $Re$ values. 
% The projection onto the modes in the latent space revealed a clear separation of modes across dynamical processes and dissipation rates. 
% Two-dimensional visualizations of the latent representation of flow states revealed new understanding of fundamental physical processes in the physical space.
% A separation of the dynamical processes towards a bursting event and away from a bursting event was learned in the latent representation at $Re = 40$, from which a particular physical structure was discovered to be central to mediating bursting dynamics.

Two-dimensional visualizations of large turbulent datasets revealed distinct clusters corresponding to different `classes' of vorticity snapshots. 
At $Re=40$, multiple types of high-dissipation events were identified.
Visualizations of finely resolved bursting trajectories highlighted a common route to a bursting event involving a large projection onto a particular dominant latent mode, and a common route back to quiescent flow through one of the high dissipation clusters. 

% Short-period orbits were shown to be mostly detached from the turbulent attractor in the low-dimensional visualizations. 
% This was particularly stark at $Re = 100$, where a lack of long-period orbits have been computed to date. 
% This apparent dynamical irrelevance of these short-period orbits was attributed to a weaker projection onto the $\ell = 0$ latent mode which spans the mean-flow state, when compared to states sampled from the turbulent attractor. 
% This finding motivates the development of new approaches to compute longer-period orbits. 
The low-dimensional visualizations of turbulence were then used to assess the coverage of the turbulent dynamics by large libraries of unstable periodic orbits at both $Re=40$ and $Re=100$.
There was a reasonable representation -- with notable gaps -- at the lower $Re$ value.
However, the visualizations at $Re=100$ revealed that the latent representations of nearly \emph{all} periodic solutions were distinct from the turbulence in a fundamental way.
An analysis of individual latent structures indicated a much weaker projection onto a structure associated with the turbulent mean profile. 
These distinctions are not apparent in simpler low-dimensional projections (e.g. production/dissipation) and indicate that these solutions may not be embedded in the turbulent attractor. 

% paragraph looking ahead to synchronization 

% One caveat of using autoencoders to estimate dimensionality is that the model attempts to
Autoencoders seek to exactly reconstruct their inputs, and hence many of the $O(Re^{1/3})$ dimensions found here may not correspond directly to dynamical variables. 
For instance, it is known in three-dimensional turbulence that scales smaller than $\sim 5\pi \eta$ (where $\eta$ is the Kolmogorov length) are `slaved' to the larger scales within the turbulent attractor \cite{lalescu2013,Zaki2024}. 
The equivalent result in two-dimensional turbulence is unknown at present, but having a model learn only the `dynamical' components of vorticity, determined via an appropriate low-pass filter, would presumably simplify the reconstruction problem and allow for a more robust estimate of the $Re$-dependence.
We hope to report on this in the near future.

\appendix

\section{Architecture details}
\label{app:arch_details}

Full details of the architecture used in this work are presented in this section. 
Dense blocks \citep{Huang2016, Huang2019} are used here in place of traditional single convolutional layers in both the encoder and decoder, similar to the approach in \cite{Page2024}. 
The blocks are groups of convolutional layers where the output of each convolutional layer is concatenated with its input. 
Each dense block consists of three convolutional layers with 32 filters each.
Periodic padding is also applied to each image before a convolution, so that the output of a convolutional layer has the same shape as the input. 
The `GELU' activation function \citep{hendrycks2023} and batch normalization \citep{ioffe15} are applied to the output of convolutional layers. 
Max pooling layers are used to gradually reduce the dimensionality of the image throughout the encoder, while upsampling layers are applied within the decoder to return the original shape at the decoder output.

\newcommand{\pc}[2]{\text{PC}(#1 \times #1, #2)}
\newcommand{\lpc}[2]{\text{LPC}(#1 \times #1, #2)}
\newcommand{\db}[1]{\text{DB}(#1 \times #1)}
\newcommand{\p}[1]{\text{MP}(#1 \times #1)}
\newcommand{\up}[1]{\text{UP}(#1 \times #1)}

The encoder architecture can be summarized by the following sequence of operations:
\begin{align}
    \omega &\to \pc{8}{32} \to \db{8} \to \p{2} \nonumber\\
    &\to \pc{4}{32} \to \db{4} \to \p{2} \nonumber\\
    &\to \pc{4}{32} \to \db{4} \to \p{2}  \nonumber\\
    &\to \pc{4}{32} \to \db{4} \nonumber \\
    &\to \pc{4}{4} \to \text{Flatten} = \mathscr{E}(\omega),
\end{align}
where the first term in the brackets denotes the size of the convolutional filters and the second term denotes the number of filters, `PC' stands for a convolutional layer with periodic padding, `DB' for the dense blocks described above and `MP' for a max pooling layer. 
The input vorticity field is a single-channel image of size $128^2$, hence the output of the encoder $\mathscr{E}(\omega)$ is a flattened vector of length $16^2 \times 4 = 1024$. 

The vector $\mathscr E(\omega)$ is then passed through a series of four fully connected (`FC') linear layers with equal input/output dimension. 
These layers do not include `bias' terms and hence are pure matrix multiplications of the upstream input.
The architecture of the bottleneck is thus summarized by
\begin{equation}
    \mathscr{E}(\omega) \to FC(1024)^4 = \mathscr{W}(\mathscr{E}(\omega)).
\end{equation}
This output $\mathscr{W}(\mathscr{E}(\omega))$ is the latent representation learned by the autoencoder. %, which is driven to be of minimal rank due to the implicit rank minimization of the four fully connected linear layers. 
The impact of the internal linear operations is to drive the representation to low rank, something which has also been empirically observed in other studies \cite{jing2020,zeng2021}.

To decode this latent representation back to physical space, $\mathscr{W}(\mathscr{E}(\omega))$ is reshaped into an `image' with four channels of dimension $16^2$.
The structure of the decoder used then essentially mirrors the encoder (replacing max pooling layers with upsampling layers), with the exception of the addition of an output stream of linear periodic convolutional layers (`LPC'):
\begin{align}
    \mathscr{W}(\mathscr{E}(\omega)) &\to \text{Reshape} \nonumber\\
    &\to \pc{4}{32} \to \db{4} \to \up{2} \nonumber\\
    &\to \pc{4}{32} \to \db{4} \to \up{2} \nonumber\\
    &\to \pc{4}{32} \to \db{4} \to \up{2}  \nonumber\\
    &\to \pc{8}{32} \to \db{8} \nonumber \\
    &\to \lpc{16}{1} \to \lpc{12}{1} \nonumber\\
    &\to \lpc{8}{1} \to \lpc{6}{1} \nonumber\\ 
    &= \mathscr{D}(\mathscr{W}(\mathscr{E}(\omega))) = \mathscr{A}(\omega).
\end{align}
These linear layers result in a more faithful reproduction of the spectral content of the input snapshots (see \S\ref{sec:network}).
%The motivation for this final stream of linear convolutional layers with decreasing filter size is to encourage the network to accommodate for the multi-scale nature of the input data. 
%The energy spectrum of the reconstructed vorticity fields was computed using this output stream and using a single $\lpc{8}{1}$ layer.
%As presented in the main text, the spectrum using the output stream was found to match the true spectrum of the data up to higher wavenumbers, when compared to the spectrum using the single linear layer.

% \section{Modal structures at $Re = 200, 400$}
% \label{app:modal_highRe}

\section{Unstable eigenfunctions}
\label{app:unstable}

\begin{figure}
    \centering
    \includegraphics[width=\linewidth]{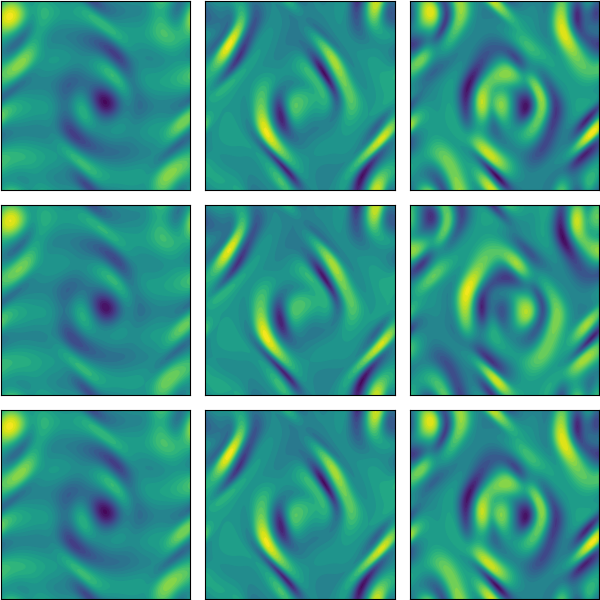}
    \caption{Out-of-plane vorticity for three low-dissipation exact solutions at $Re=40$ (left), alongside the real (middle) and imaginary (right) components of the leading unstable eigenfunction. (Top) A traveling wave solution with a wavespeed of 0.0198. (Middle) A UPO with period $T = 12.207$ and translational shift of $s = 0.243$. (Bottom) A UPO with $T = 5.81$ and $s = 0.115$.}
    \label{fig:TW_stab}
\end{figure}
We include here three examples of low-dissipation exact solutions at $Re=40$, along with a visualization of their leading unstable eigenfunctions. 
These snapshots are reported in figure \ref{fig:TW_stab} and should be compared with decodes of the leading SVD modes reported in figure \ref{fig:modes}.

\section{Hyperparameter sensitivity}
\label{app:hyperparam}

\AC{Hyperparameter sensitivity for the IRMAE networks was checked using a smaller (2 convolutional layers per dense block, smaller embedding dimension of $16^2 =256$) version of the final network outlined in appendix \ref{app:arch_details}. 
A smaller network was used to perform this hyperparameter analysis as the full IRMAE networks required $\sim 2$ weeks of wall-time, due to the large datasets and small batch sizes considered.
The average and standard deviation of the relative error is reported in table \ref{tab:learningrate_results} for a number of combinations of batch size and initial learning rate.}

\begin{table}%[H] add [H] placement to break table across pages
\caption{\AC{Sensitivity of the average $\varepsilon$ and standard deviation $\sigma$ of the relative error on the initial learning rate and batch size.} \label{tab:learningrate_results}}
\begin{ruledtabular}
\begin{tabular}{c|c|c|c}
Initial learning rate & Batch size  & $\varepsilon$ & $\sigma$  \\ \hline
        $10^{-4}$ & 16 & 0.021 & 0.015 \\
        $5\times 10^{-3}$ & 16 & 0.032 & 0.0196 \\
        $10^{-3}$ & 16 &  0.038  &  0.025 \\
        $10^{-4}$ & 8 & 0.0186 & 0.0116 \\
        $10^{-4}$ & 16 & 0.0187 & 0.0141 \\
        $10^{-4}$ & 64 & 0.0188  &  0.0133
\end{tabular}
\end{ruledtabular}
\end{table}

\begin{figure}
    \centering
    \includegraphics[width=\linewidth]{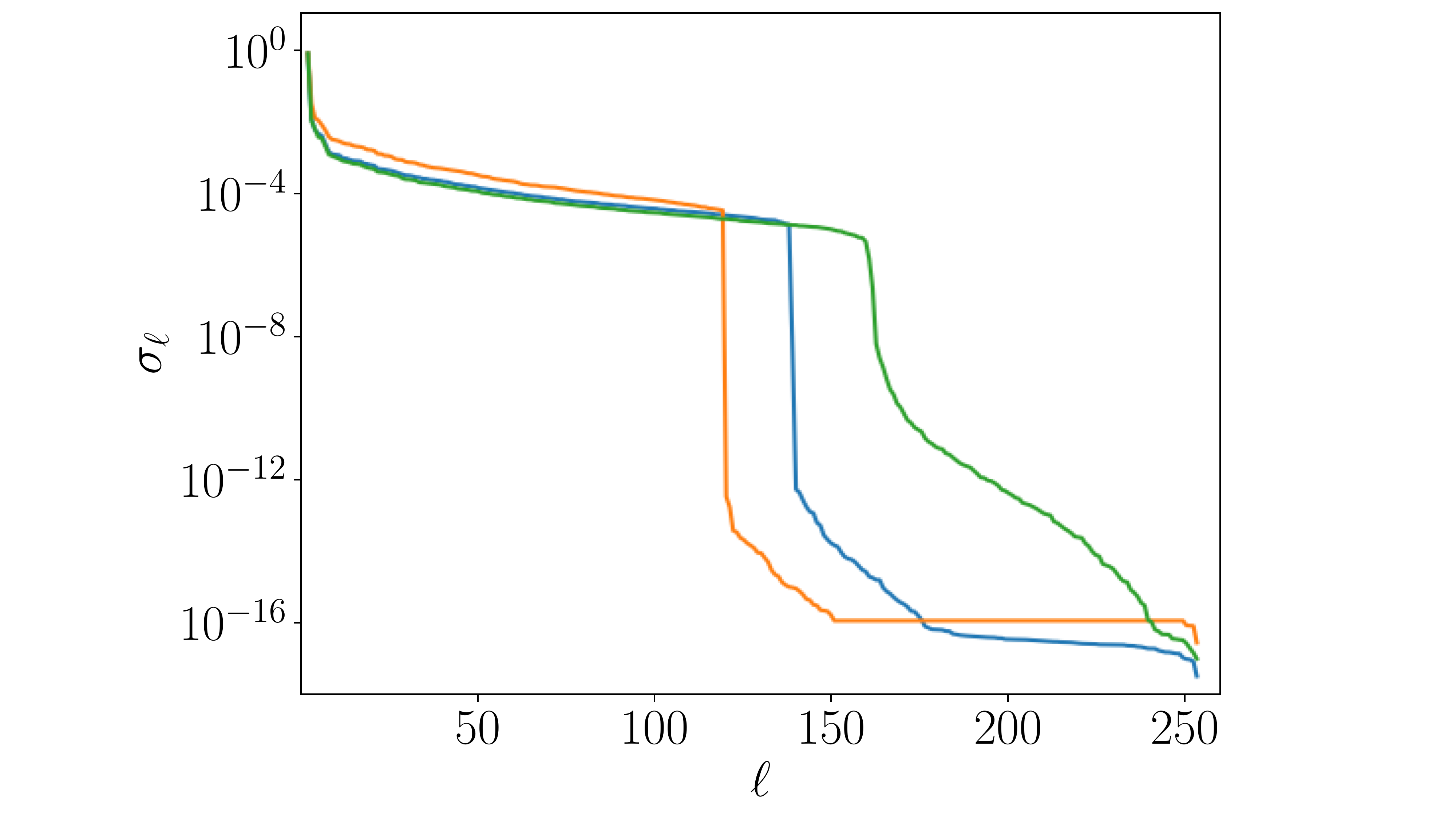}
    \caption{\AC{Singular values $\sigma_{\ell}$ in the singular value decomposition of the covariance of the latent data matrix for a smaller network trained with a batch size of 64 (green), 16 (blue) and 8 (orange).}}
    \label{fig:batchsize_svd}
\end{figure}

\AC{An initial learning rate of $10^{-4}$ yields the best relative error, while the relative error is quite robust to changes in the batch size. 
However, the rank of the learned representation is significantly smaller for smaller batch sizes, as shown in figure \ref{fig:batchsize_svd}.
Keeping the size of the training dataset and the number of epochs constant, a smaller batch size implies that a greater number of gradient descent steps are taken. 
This result is supportive of the idea that each gradient update of the weights incrementally drives the rank of the learned latent space downwards.}

\begin{acknowledgments}

This research has been supported by the UK Engineering and Physical Sciences Research Council through the MAC-MIGS Centre for Doctoral Training (EP/S023291/1). 
Computations were performed on  the Cirrus UK National Tier-2 HPC Service at EPCC (http://www.cirrus.ac.uk).
JP acknowledges support from UKRI Frontier Guarantee Grant EP/Y004094/1.
We are grateful to Masanobu Inubushi for pointing us to reference \cite{Liu1993} with a lower bound for the dimension of the global attractor.
JP also thanks Rich Kerswell for useful discussions and Mike Graham for helpful discussion on the IRMAE method.

\end{acknowledgments}

\bibliography{bib}

\end{document}